\documentclass[11pt,letterpaper]{article}
\pdfoutput=1
\usepackage{jcappub}

\usepackage{appendix}
\usepackage{array}
\newcolumntype{x}[1]{>{\centering\arraybackslash}p{#1}}

\usepackage{amsmath}
\usepackage{amsfonts}
\usepackage{amssymb}
\usepackage{eucal}               
\usepackage{mathrsfs}          
\usepackage{slashed}

\usepackage{epsfig}
\usepackage{graphicx}
\usepackage{dcolumn}
\usepackage{enumerate}

\usepackage{color}
\usepackage{ulem}
\hypersetup{colorlinks=false}

\newcommand{\virg}[1]{`#1'}

\newcommand{\eg}{e.g.~}
\newcommand{\ie}{i.e.~}

\newcommand{\beq}{\begin{equation}}
\newcommand{\eeq}{\end{equation}}
\newcommand{\ud}{\text{d}}

\newcommand{\mDM}{m}
\newcommand{\ER}{E_\text{R}}
\newcommand{\Ed}{E'}

\newcommand{\vmin}{v_\text{min}}

\newcommand{\bol}[1]{\boldsymbol{#1}}
\newcommand{\bfv}{\bol{v}}
\newcommand{\eH}{\mathcal{H}}
\newcommand{\eR}{\mathcal{R}}

\newcommand{\Lag}{\mathscr{L}}	

\title{Direct detection of Light Anapole and Magnetic Dipole DM}

\author[a]{Eugenio Del Nobile,}
\author[a]{Graciela B. Gelmini,}
\author[b]{Paolo Gondolo,}
\author[a]{and Ji-Haeng Huh}

\affiliation[a]{Department of Physics and Astronomy, UCLA,\\
475 Portola Plaza, Los Angeles, CA 90095, USA}
\affiliation[b]{Department of Physics and Astronomy, University of Utah,\\
115 South 1400 East \#201, Salt Lake City, UT 84112, USA}

\emailAdd{delnobile@physics.ucla.edu}
\emailAdd{gelmini@physics.ucla.edu}
\emailAdd{paolo@physics.utah.edu}
\emailAdd{jhhuh@physics.ucla.edu}

\abstract{
We present comparisons of direct detection data for ``light WIMPs'' with an anapole moment interaction (ADM) and a magnetic dipole moment interaction (MDM), both assuming the Standard Halo Model (SHM) for the dark halo of our galaxy and in a halo-independent manner. In the SHM analysis we find that a combination of the 90\% CL LUX and CDMSlite limits or the new 90\% CL SuperCDMS limit by itself exclude the parameter space regions allowed by DAMA, CoGeNT and CDMS-II-Si data for both ADM and MDM. In our halo-independent analysis the new LUX bound excludes the same potential signal regions as the previous XENON100 bound. Much of the remaining signal regions is now excluded by SuperCDMS, while the CDMSlite limit is much above them. The situation is of strong tension between the positive and negative search results both for ADM and MDM. We also clarify the confusion in the literature about the ADM scattering cross section.
}


\begin{document}

\maketitle

\section{Introduction}
Some of the dark matter (DM) candidates most actively searched-for are weakly interacting massive particles (WIMPs), \ie particles with weakly interacting cross sections and masses between about $1$ GeV/$c^2$ and $10$ TeV/$c^2$. Particular attention has been given in recent times to ``light WIMPs'' with masses around $1$--$10$ GeV/$c^2$.
In 2004 they were first shown \cite{Gelmini:2004gm} to provide a viable DM interpretation of the DAMA annual modulation \cite{Bernabei:2003za} compatible with all negative searches at the time, assuming spin-independent isospin-conserving interactions with nuclei and the Standard Halo Model (SHM) for the dark halo of our galaxy.
The interest in these candidates intensified after the first DAMA/LIBRA 2008 results \cite{Bernabei:2008yi}, and even more in recent years, due to other potential DM hints in the same mass and cross section region.

At present, four direct DM detection experiments (DAMA \cite{Bernabei:2010mq}, CoGeNT \cite{Aalseth:2010vx, Aalseth:2011wp, Aalseth:2012if,Aalseth:2014eft,Aalseth:2014jpa},  CRESST-II \cite{Angloher:2011uu}, and now CDMS-II-Si \cite{Agnese:2013rvf}) have data  that may be interpreted as  signals from light WIMPs. DAMA \cite{Bernabei:2010mq} and CoGeNT \cite{Aalseth:2011wp, Aalseth:2014eft, Aalseth:2014jpa} report annual modulations in their event rates compatible with those expected for a DM signal \cite{Drukier:1986tm}. CoGeNT \cite{Aalseth:2010vx, Aalseth:2012if,Aalseth:2014eft, Aalseth:2014jpa}, CRESST-II \cite{Angloher:2011uu}, and CDMS-II-Si \cite{Agnese:2013rvf} observed an excess of events above their expected backgrounds. Other experiments however do not report any potential positive signal, thus placing upper limits on DM WIMPs. The most stringent limits on the unmodulated rate for light WIMPs come from the LUX \cite{Akerib:2013tjd}, XENON10 \cite{Angle:2011th}, XENON100 \cite{Aprile:2011hi, Aprile:2012nq}, CDMS-II-Ge \cite{Ahmed:2010wy}, CDMSlite \cite{Agnese:2013jaa} and SuperCDMS \cite{Agnese:2014aze} experiments, with the addition of SIMPLE \cite{Felizardo:2011uw}, PICASSO \cite{Archambault:2012pm} and COUPP \cite{Behnke:2012ys} for spin-independent isospin-violating \cite{Kurylov:2003ra, Feng:2011vu} and spin-dependent interactions. CDMS-II-Ge \cite{Ahmed:2012vq} negative search for an annual modulation in their data constrains directly the amplitude of a hypothetical annually modulated signal.

We have so far mentioned only potential contact interactions. However, an interesting possibility explored recently  is that of neutral DM particle candidates which interact with photons through higher electromagnetic moments. For example, these can occur if the DM particle couples at tree level to heavier charged states, and thus to the photon at loop level, as it happens with neutrinos in the Standard Model. Another possibility is that the neutral DM particle is a composite object made of charged particles, as a ``dark neutron''. In these cases, the interaction at low energy is described by non-renormalizable effective operators suppressed by the scale of new physics, expected to be proportional to the mass of the particles running in the loop, or otherwise, to the compositeness scale of the ultraviolet theory.

For fermionic DM, the most studied candidates with electromagnetic moments are WIMPs with magnetic or electric dipole moments, \eg in Refs.~\cite{Pospelov:2000bq, An:2010kc, Sigurdson:2004zp, Barger:2010gv, Chang:2010en, Cho:2010br, Heo:2009vt, Gardner:2008yn, Masso:2009mu, Banks:2010eh, Fortin:2011hv, Kumar:2011iy, Barger:2012pf, DelNobile:2012tx, Cline:2012is, Weiner:2012cb, Tulin:2012uq, Cline:2012bz}, which have the lowest order electromagnetic interactions possible, given by dimension five effective operators. Magnetic and electric dipole moments vanish for Majorana fermions (although nonzero transition moments are possible), and the only possible electromagnetic moment is an anapole moment. The interaction in this case is described by an effective dimension six operator.

In particular, anapole moment DM (ADM) has been studied in the context of direct detection \cite{Pospelov:2000bq, Ho:2012bg, Fitzpatrick:2010br, Frandsen:2013cna, Gresham:2013mua} and colliders \cite{Gao:2013vfa}. Assuming the SHM, Ref.~\cite{Frandsen:2013cna} found that ADM could explain the recent CDMS-II-Si excess while still being compatible with the XENON10 and XENON100 null results, for WIMPs with mass $m\simeq 7$ GeV/$c^2$. In Ref.~\cite{Gresham:2013mua}, the CDMS-II-Si allowed region in the ADM parameter space assuming the SHM was found to be in tension with the LUX bound, although WIMPs with masses $m \simeq 7$ GeV/$c^2$ were found to be compatible with the LUX null results when a more conservative choice than that adopted by the XENON100 and LUX collaborations was used for the effective luminosity.

Here we consider ADM and MDM as potential explanations of the signal found in direct detection data, both assuming the SHM, and in a halo-independent analysis \cite{DelNobile:2013cva}. The halo-independent analysis was proposed in Refs.~\cite{Fox:2010bz, Frandsen:2011gi, Gondolo:2012rs, Frandsen:2013cna, DelNobile:2013cta, DelNobile:2013gba} (see also \cite{HerreroGarcia:2011aa, HerreroGarcia:2012fu, Bozorgnia:2013hsa}) and recently generalized in \cite{DelNobile:2013cva} to be applied to WIMP scattering cross sections with any type of speed dependency. This method consists in mapping the rate measurements and bounds onto $\vmin$ space, where $\vmin$ is the minimum WIMP speed necessary to impart to the target nucleus a recoil energy $\ER$. This method allows to factor out a function of $\vmin$ which gives the dependency of the rate on the DM velocity distribution, and use this as a detector independent variable. Since $\vmin$ is also a detector-independent quantity, while $\ER$ is not (it depends on the target mass), outcomes from different direct detection experiments can be directly compared in $\vmin$ space.

\section{Data analysis}

The data analysis in this paper is the same as in Ref.~\cite{DelNobile:2013gba}, except for the recent CoGeNT 2014 data \cite{Aalseth:2014eft, Aalseth:2014jpa} and SuperCDMS \cite{Agnese:2014aze} data.

For SuperCDMS, we compute the $90\%$ CL bound with the Maximum Gap method \cite{Yellin:2002xd} in the $1.6$-$10$ keVnr energy range, considering the eleven candidate events found in Ref.~\cite{Agnese:2014aze} with an exposure of 577 kg days. For the efficiency we use the red line in Fig.~1 of Ref.~\cite{Agnese:2014aze}.

The CoGeNT 2014 \cite{Aalseth:2014eft} region and bound in our SHM analysis is obtained using only the unmodulated rate. Adding the modulation data does not change the region significantly. In our halo-independent analysis we use both the rate and modulation amplitude data \cite{Aalseth:2014eft}.

\begin{figure}[t]
\centering
\includegraphics[width=0.49\textwidth]{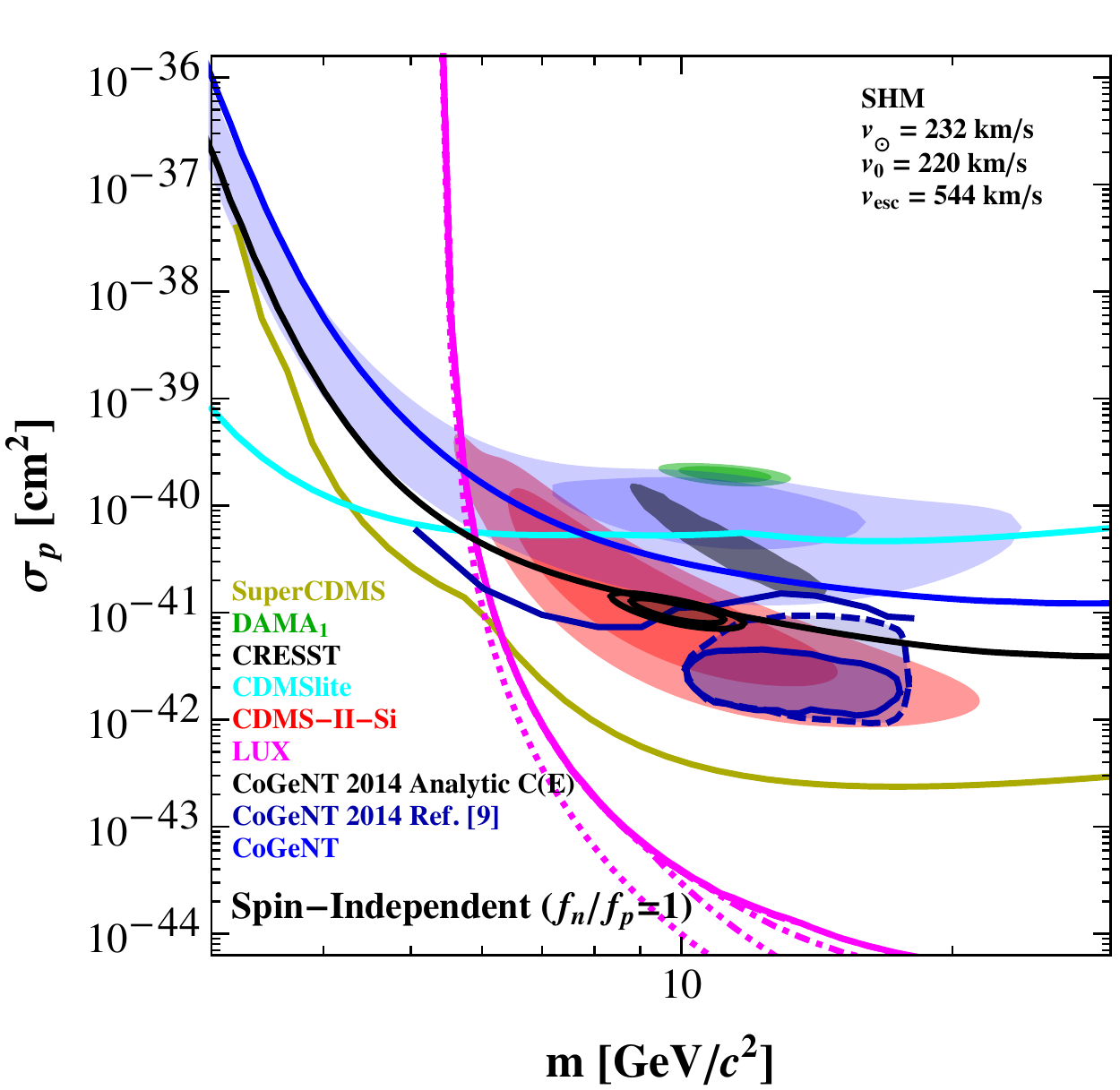}
\includegraphics[width=0.49\textwidth]{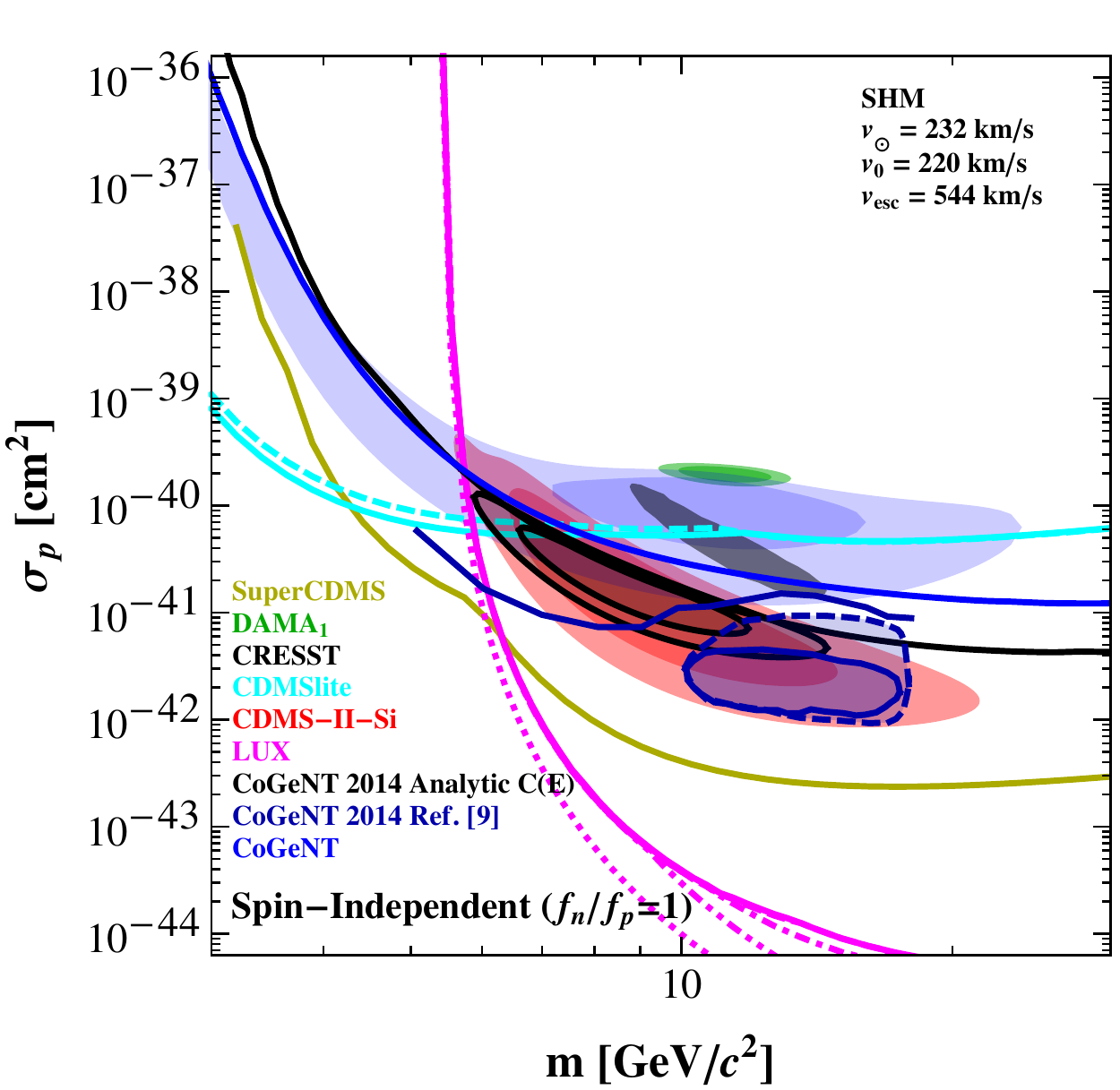}
\caption{\label{fig:msigmaSI}
$90\%$ CL bounds and $68\%$ and $90\%$ CL allowed regions in the spin-independent isospin-conserving WIMP proton cross section, $\sigma_p$, vs WIMP mass plane, assuming the SHM (see section \ref{SHM}). The CRESST-II low mass allowed region, taken from Ref.~\cite{Angloher:2011uu}, is only shown at the $2 \sigma$ CL in dark gray. The dark blue \virg{CoGeNT 2014 Ref.~[9]} region and limit are taken from Ref.~\cite{Aalseth:2014jpa}. The CoGeNT 2011-2012 modulation data yield the blue regions. In Fig.~\ref{fig:msigmaSI}a (left panel) the black contour CoGeNT 2014 region corresponds to using the analytic form for $C(E)$ while Fig.~\ref{fig:msigmaSI}b (right panel) shows the same region but with $C(E)$ taken from the data points of Fig.~21 of Ref.~\cite{Aalseth:2012if}. Fig.~\ref{fig:msigmaSI}b also shows the negligible difference between the CDMSlite bounds obtained as in Ref.~\cite{DelNobile:2013gba} (solid cyan line) and using their now publicly available data table \cite{ref:CDMSlite} (dashed cyan line).
}
\end{figure}

We analyze the CoGeNT 2014 data \cite{Aalseth:2014eft} in the same way we analyze the older 2011-2012 data \cite{Aalseth:2010vx, Aalseth:2011wp, Aalseth:2012if} following the procedure specified in Ref.~\cite{Aalseth:2012if}, but with a different choice of the residual surface event correction $C(E)$. Using the analytic expression $C(E) = 1 - \exp(-aE)$ with $E$ in keVee and $a = 1.21 \pm 0.11$ we find that the errors in the resulting bulk rate are underestimated. Thus we use the data points for $C(E)$ in Fig.~21 of Ref.~\cite{Aalseth:2012if} with their corresponding error bars. As we can see in Fig.~\ref{fig:msigmaSI}, the CoGeNT 2014 region using the analytic $C(E)$ form (labeled \virg{CoGeNT 2014 Analytic $C(E)$} in Fig.~\ref{fig:msigmaSI}a) is much smaller than the region obtained by using the 12 data points of Fig.~21 of \cite{Aalseth:2012if} (shown in Fig.~\ref{fig:msigmaSI}b and labeled \virg{CoGeNT 2014}). Fig.~\ref{fig:msigmaSI} shows the regions and bounds in the usual WIMP-proton cross section $\sigma_p$ vs WIMP mass $m$ plane assuming the SHM for WIMPs with spin-independent isospin-conserving interactions (\ie with equal couplings to neutrons and protons, $f_n = f_p$). For comparison we include the CoGeNT 2014 regions and limit taken from Ref.~\cite{Aalseth:2014jpa}. These regions (shown in dark blue) are shifted toward larger masses and smaller cross sections respect to the regions we obtain, but the two are compatible within the error bars.

Although the $C(E)$ in Fig.~21 of \cite{Aalseth:2012if} was derived only using the CoGeNT 2011-2012 data, we think it is still reasonable to use it for the cumulative 2014 data. In particular the $C(E)$ value derived from the two log-normal curves in Fig.~1a of \cite{Aalseth:2014jpa} for the 0.5 to 0.7 keVee energy interval is $C(E) = 0.49$, entirely compatible with the $C(E)$ in the same energy interval in Fig.~21 of \cite{Aalseth:2012if}, $C(E) = 0.47 \pm 0.34$.

Notice that in Figs.~\ref{fig:msigmaSI} and \ref{fig:msigma} we also include the CoGeNT 2011-2012 $m$-$\sigma$ regions derived from the annual modulation (only labeled \virg{CoGeNT} and shown in blue) centered at larger cross sections but still compatible within error bars.

For the halo-independent analysis of the CoGeNT 2014 unmodulated rate data we apply the $C(E)$ in the 12 energy bins of Fig.~21 of Ref.~\cite{Aalseth:2012if} and then combine the resulting bulk rate into only four energy bins, 0.5 to 1.1 keVee, 1.1 to 1.7 keVee, 1.7 to 2.3 keVee and 2.3 to 2.9 keVee. For the modulation data, the $C(E)$ correction is not needed, since it is assumed that the surface events are not annually modulated. With our choice of modulation phase (DAMA's best fit value of 152.4 days from January $1^\text{st}$) and modulation period of one year in our fit of the data, the modulation amplitude we find in the first and fourth energy bins are negative, thus we show their modulus in the corresponding figures with thinner solid lines. However, in both bins the modulation amplitude is compatible with zero, at the $0.7 \sigma$ and $1.8 \sigma$ confidence level for the first and fourth, respectively. Also in the second and third energy bins the amplitudes are compatible with zero, at the $1.3 \sigma$ and $0.8 \sigma$ CL respectively.

\section{Anapole moment dark matter}
The possibility that particles could have an anapole moment was first proposed by Ya. Zel'dovich in 1957 \cite{ZelDovich}.
The anapole moment interaction breaks C and P, but preserves CP. It was first measured experimentally in cesium atoms in 1997 \cite{Wood:1997zq}. The interaction with photons of a spin-$\frac{1}{2}$ Majorana fermion $\chi$ due to its anapole moment can be expressed in a Lorentz-invariant form as
\beq\label{eq:ADMlag}
\Lag = \frac{1}{2} \frac{g}{\Lambda^2} \, \bar{\chi} \gamma^{\mu} \gamma^5 \chi \, \partial^\nu F_{\mu\nu} \ ,
\eeq
where $g$ is a dimensionless coupling constant and $\Lambda$ is the new physics mass scale. In the non-relativistic limit, this contains the interaction of the particle spin with the curl of the magnetic field, $H = - (g / \Lambda^2) \, \bol{\sigma} \cdot \bol{\nabla} \times \bol{B}$. Notice that $\bar{\chi} \gamma^{\mu} \chi$ vanishes for Majorana fermions.

In Appendix \ref{xsecADM} we derive the differential cross section for scattering of an ADM Majorana fermion with mass $m$ and speed $v = | \bfv |$ with a target nucleus at rest. We find
\beq
\label{eq:ADMxsec}
\frac{\ud\sigma_T}{\ud\ER} = \sigma^A_{\rm ref} \frac{m_T}{\mu_N^2} \frac{\vmin^2}{v^2} \left[ Z^2 \left( \frac{v^2}{\vmin^2} - 1 \right) F_{{\rm E}, T}^2({\bf q}^2) + 2 \frac{\lambda_T^2}{\lambda_N^2} \frac{\mu_T^2}{m_N^2} \left( \frac{J_T + 1}{3 J_T} \right) F_{{\rm M}, T}^2({\bf q}^2) \right] \ ,
\eeq
where the square of the momentum transfer is ${\bf q}^2 = 2 m_T \ER$ with $m_T$ the nuclear mass, $m_N$ is the nucleon mass, $\mu_N$ and $\mu_T$ are the DM-nucleon and DM-nucleus reduced mass, respectively, $\lambda_N = e / 2 m_N$ is the nuclear magneton, $\lambda_T$ is the nuclear magnetic moment, $J_T$ is the nuclear spin, and we defined the ADM reference cross section
\beq
\label{eq:ADMsigref}
\sigma^A_{\rm ref} \equiv 2 \mu_N^2 \, \alpha g^2 / \Lambda^4 \ ,
\eeq
where $\alpha = e^2 / 4 \pi \simeq 1/137$ is the fine structure constant. In an elastic collision $v_{\rm min}$ is
\beq
\label{eq:vmin}
\vmin = \frac{| {\bf q} |}{2 \mu_T} = \sqrt{\frac{m_T \ER}{2 \mu_T^2}} \ .
\eeq

The first term in Eq.~\eqref{eq:ADMxsec} corresponds to the interaction with the nuclear charge $Z$ and the second term with the nuclear magnetic field. Notice that Eq.~\eqref{eq:ADMxsec} differs from some of the expressions given for the same cross section in the literature \cite{Ho:2012bg,Frandsen:2013cna}.
The equation in Refs.~\cite{Ho:2012bg} and \cite{Frandsen:2013cna} does not make the distinction between the electric and magnetic terms in the cross section and treats instead both as proportional to the nucleus charge squared. Eq.~\eqref{eq:ADMxsec} agrees with the cross section given in Refs.~\cite{Fitzpatrick:2010br} and \cite{Gresham:2013mua} except that we do not make the approximation of taking $F_{{\rm M},T}^2({\bf q}^2)$ equals to one.

The form factors we use are the standard longitudinal and transverse nuclear form factors (see e.g. Eq.~(1.29) of Ref.~\cite{Donnelly:1975ze} or Eqs.~(4.32) and (4.33) of Ref.~\cite{Donnelly:1984rg}), $F_{\rm L}^2({\bf q}^2)$ and $F_{\rm T}^2({\bf q}^2)$, but normalized to 1 at ${\bf q}^2=0$. Thus, using Eqs.~(4.44), (4.48) and (4.64) of Ref.~\cite{Donnelly:1975ze} we define the electric and magnetic form factors as
\begin{eqnarray}
\label{eq:FE}
F^2_{\rm E}({\bf q}^2)&=&\frac{4\pi}{Z^2}F_{\rm L}^2({\bf q}^2)\ ,\\
\label{eq:FM}
F^2_{\rm M}({\bf q}^2)&=&\frac{3J_T}{J_T+1}\frac{8\pi m_N^2}{{\bf q}^2}\frac{\lambda_N^2}{\lambda_T^2}F_{\rm T}^2({\bf q}^2)\ .
\end{eqnarray}
In terms of the Fourier transforms $\rho({\bf q})$ of the electromagnetic charge density and $J_\lambda({\bf q})$ with $\lambda=\pm 1$ of the transverse component of the current density, the longitudinal and transverse form factors are defined as (see e.g. Eqs.~(1.22) and (1.28) of Ref.~\cite{Donnelly:1975ze})
\begin{eqnarray}
F^2_{\rm L}({\bf q}^2)&=&\frac{1}{4\pi(2J_T+1)}\sum_{M_i}\sum_{M_f}|\rho({\bf q})|^2\\
F^2_{\rm T}({\bf q}^2)&=&\frac{1}{4\pi(2J_T+1)}\sum_{M_i}\sum_{M_f}J_\lambda({\bf q}){J_\lambda}^*({\bf q}).
\end{eqnarray}
Here $M_i$ and $M_f$ are the initial and final projections of the target nuclear angular momentum $J_T$.

For small $|{\bf q}|$, the monopole component of the charge distribution gives the dominant contribution to the electric form factor $F_{\rm E}^2({\bf q}^2)$ (see Eqs.~(4.42), (4.52) and (4.53) of Ref.~\cite{Donnelly:1984rg}) which we take to be the Helm form factor \cite{Helm:1956zz}. The magnetic form factor $F_{\rm M}({\bf q}^2)$ has contributions from the magnetic moments of the nucleons as well as from the magnetic currents due to the orbital motion of the protons.

For the light WIMPs we consider in the following, the second term in Eq.~\eqref{eq:ADMxsec} is negligible for all the target nuclei we deal with except sodium. This term is more important for lighter nuclei, such as sodium and silicon, but silicon has a very small magnetic dipole moment. The nuclear magnetic moment of $^{23}{\rm Na}$ is $\lambda_{\rm Na}/\lambda_N=2.218$. We took the magnetic form factor $F_{\rm M}^2({\bf q}^2)$ from Fig.~31 of Ref.~\cite{Donnelly:1984rg} which is fitted well by the approximate functional form $F_{\rm M,Na}^2({\bf q}^2)=(1-1.15845{\bf q}^2+0.903442{\bf q}^4)\exp(-2.30722{\bf q}^2)$, where $|{\bf q}|$ is in units of fm$^{-1}$.

One can immediately notice that the velocity dependence of the differential cross section in Eq.~\eqref{eq:ADMxsec} is different from the dependence of the cross section in the usual case of contact spin-independent interaction, which is $\ud\sigma_T / \ud\ER \propto 1/v^2$. What is very important for the application of the halo-independent formalism is the presence of two terms with different velocity dependence, which makes it necessary to resort to the generalization of the method, presented in Ref.~\cite{DelNobile:2013cva}. There, we studied DM with magnetic dipole moment (MDM) whose cross section
\beq
\label{eq:MDMxsec}
\frac{\ud\sigma_T}{\ud\ER} = \sigma^M_{\rm ref} \frac{m_T}{\mu_T^2} \frac{1}{v^2} \left[ Z^2 \left( \frac{v^2}{\vmin^2} - 1 + \frac{\mu_T^2}{m^2}\right) F_{{\rm E}, T}^2({\bf q}^2) + 2 \frac{\lambda_T^2}{\lambda_N^2} \frac{\mu_T^2}{m_N^2} \left( \frac{J_T + 1}{3 J_T} \right) F_{{\rm M}, T}^2({\bf q}^2) \right] \ ,
\eeq
also has two terms with different velocity dependence. The MDM reference cross section is defined as
\beq
\label{eq:MDMsigref}
\sigma_{\rm ref}^M \equiv \alpha \lambda_\chi^2\ ,
\eeq
with $\lambda_\chi$ the DM magnetic dipole moment.

\section{Halo-dependent analysis}
\label{SHM}

For this analysis we use the SHM with a truncated Maxwell-Boltzmann velocity distribution in the galactic reference frame \cite{Savage:2008er},
\beq
f_G(\bol{u})=\frac{e^{-\bol{u}^2/v_0^2}}{(v_0\sqrt{\pi})^3N_{\rm esc}}\Theta(v_{\rm esc}-|\bol{u}|)
\eeq
normalized to 1. Here $\bol{u}=\bfv+\bfv_{\rm E}(t)$ where $\bfv$ is the velocity of the WIMP with respect to Earth and $\bfv_{\rm E}$ is the velocity of Earth with respect to the galaxy, whose average value is $\bfv_\odot$. We use $|\bfv_\odot|=232$ km/s \cite{Savage:2008er}, ${v}_0=220$ km/s, and ${v}_{\rm esc}=544$ km/s \cite{Smith:2006ym}
. For the local DM density $\rho$ we use 0.3 GeV/$c^2$/cm$^3$.

In direct DM detection searches, the primary observable is the scattering rate within a detected energy interval $E' \in [\Ed_1, \Ed_2]$,
\begin{multline}
\label{eq:R}
R_{[\Ed_1, \Ed_2]}(t) = \frac{\rho}{\mDM} \sum_T \frac{C_T}{m_T} \int_0^\infty \ud \ER \, \int_{v \geqslant v_\text{min}(\ER)} \hspace{-18pt} \ud^3 v \, f(\bfv, t) \, v \, \frac{\ud \sigma_T}{\ud \ER}(\ER, v)
\\
\times
 \int_{\Ed_1}^{\Ed_2} \ud\Ed \, \epsilon(\Ed) G_T(\ER, \Ed) \ ,
\end{multline}
(see \eg Ref.~\cite{DelNobile:2013cva} for details). Here $C_T$ is the mass fraction of the target $T$, $f(\bfv, t)$ is the DM velocity distribution in the reference frame of Earth, and $\epsilon(\Ed)$ is the experimental acceptance. $G_T(\ER, \Ed)$ is the detector target-dependent resolution function, expressing the probability that a recoil energy $\ER$ is measured as $\Ed$, and incorporates the mean value $\langle \Ed \rangle = Q_T(\ER) \ER$, with $Q_T$ the quenching factor, and the detector energy resolution. 

Once the halo model is determined the rate depends only on the WIMP mass and a reference cross section $\sigma_{\rm ref}$ defined in Eq.~\eqref{eq:ADMsigref} for ADM and in Eq.~\eqref{eq:MDMsigref} for MDM.
\begin{figure}[t]
\centering
\includegraphics[width=0.49\textwidth]{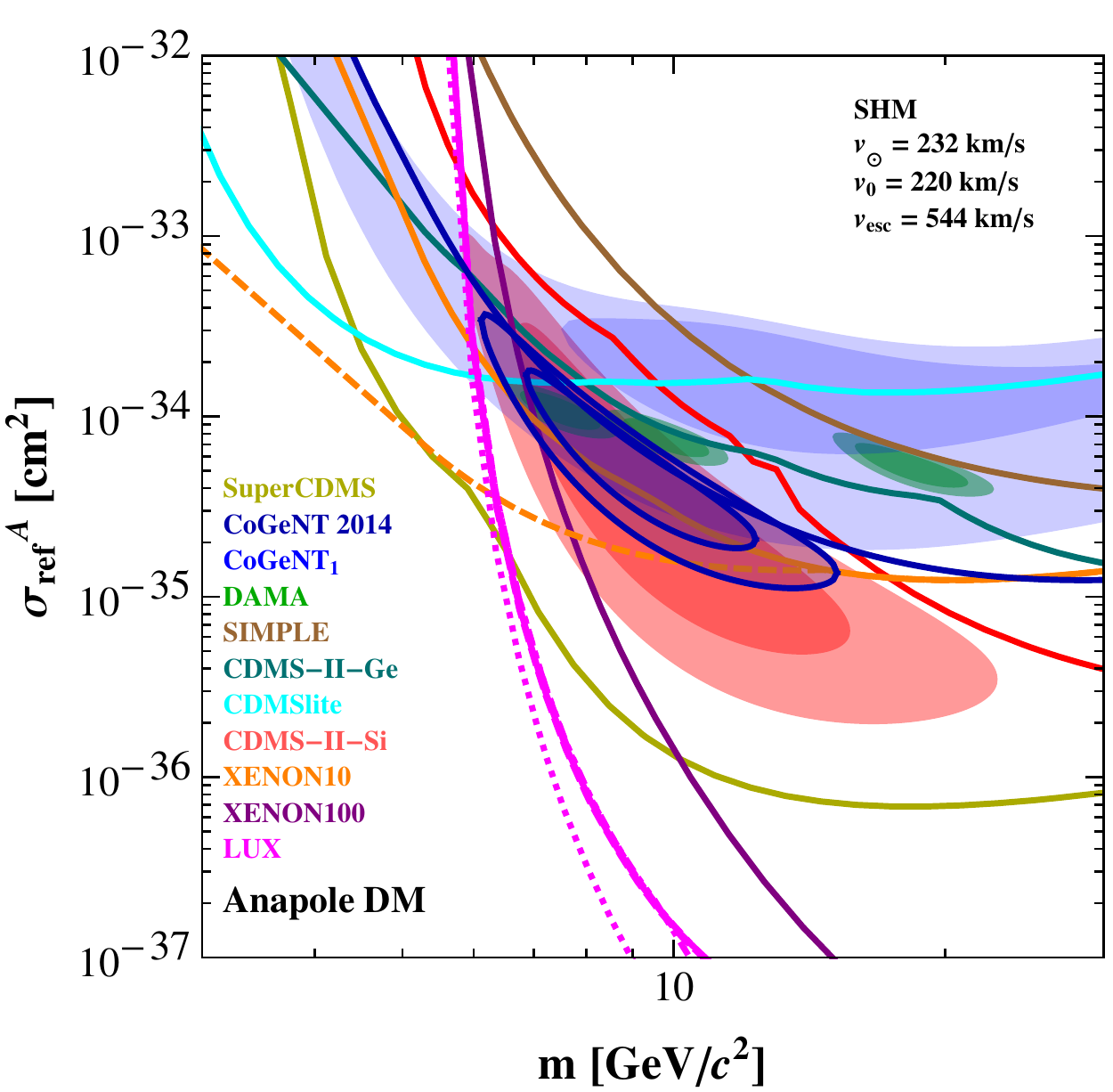}
\includegraphics[width=0.49\textwidth]{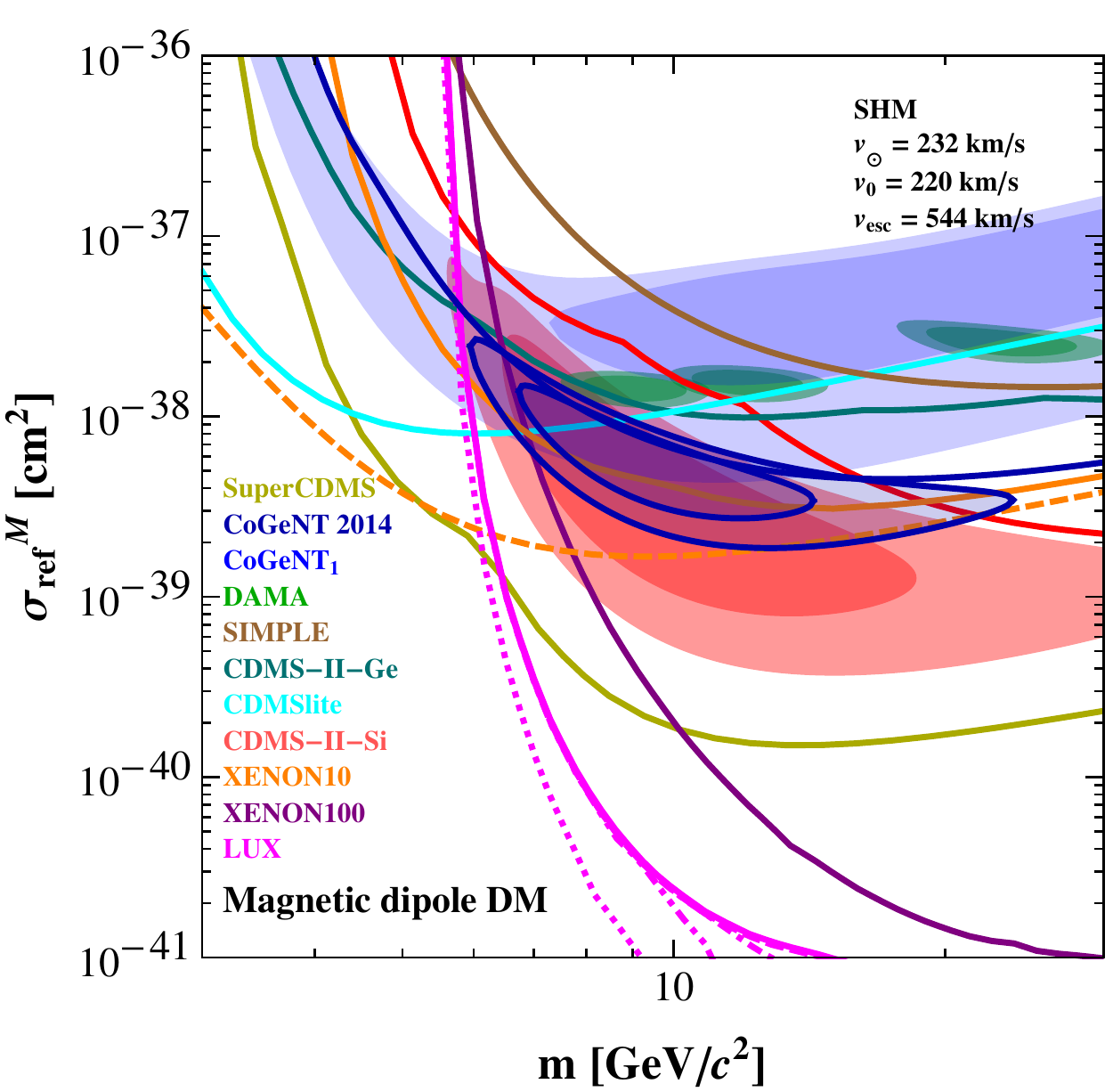}
\caption{\label{fig:msigma}
$90\%$ CL bounds from SIMPLE, CDMS-II-Ge low threshold, CDMS-II-Si, CDMSlite, SuperCDMS, XENON10, XENON100, LUX and CoGeNT 2014 data and $68\%$ and $90\%$ CL allowed regions for the DAMA and CoGeNT 2011-2012 modulation data (indicated by a subscript 1) and CoGeNT 2014 and CDMS-II-Si unmodulated data, for ADM (left) and MDM (right) assuming the SHM. The three DAMA regions correspond to $Q_{\rm Na}$ equal to $0.45$ (left), $0.30$ (middle) and the energy dependent value from Ref.~\cite{Collar:2013gu} (right).
}
\end{figure}

In Fig.~\ref{fig:msigma} we present allowed regions and upper bounds in the $m$--$\sigma_{\rm ref}$ parameter space for ADM (left panel) and MDM (right panel). We present 68\% and 90\% CL allowed regions for the DAMA \cite{Bernabei:2010mq} (sodium only) and CoGeNT 2011-2012 \cite{Aalseth:2011wp} modulation, and for the CoGeNT 2014 \cite{Aalseth:2014eft,Aalseth:2014jpa} and CDMS-II-Si \cite{Agnese:2013rvf} unmodulated rate data. For DAMA, each panel presents three regions, corresponding to three different choices of the sodium quenching factor: from left to right they are $Q_{\rm Na}=0.45$, $Q_{\rm Na}=0.30$ and the energy dependent $Q_{\rm Na}(E_R)$ from Ref.~\cite{Collar:2013gu}, which is the lowest of the three. The larger values of $Q_{\rm Na}$ yield DAMA regions which overlap more with the CDMS-II-Si and CoGeNT 2014 regions. All the allowed regions overlap in this case for WIMPs of mass 7 to 10 GeV/$c^2$ for ADM and about 8 to 10 GeV/$c^2$ for MDM, however the overlapping regions are rejected by several bounds.

In Fig.~\ref{fig:msigma} we include the 90\% CL bounds from SIMPLE \cite{Felizardo:2011uw}, CDMS-II-Ge low threshold analysis \cite{Ahmed:2010wy}, CDMS-II-Si \cite{Agnese:2013rvf}, CDMSlite \cite{Agnese:2013jaa}, SuperCDMS \cite{Agnese:2014aze}, XENON10 S2-only analysis \cite{Angle:2011th}, XENON100 \cite{Aprile:2012nq}, and LUX \cite{Akerib:2013tjd}. For XENON10 and LUX we have multiple lines. For XENON10 the orange solid line is produced by conservatively setting the electron yield ${\cal Q}_y$ to zero below 1.4 keVnr as in Ref.~\cite{Aprile:2012nq}, while for the dashed line we do not make this cut \cite{Gondolo:2012rs,DelNobile:2013cta,DelNobile:2013gba}. For LUX the different limits correspond to 0 (dotted line), 1 (double-dotted-dashed line), 3 (dotted-dashed line), 5 (dashed line) and 24 (solid line) observed events (see Ref.~\cite{DelNobile:2013gba} for a more complete description of these limits). However, in the range of masses and cross sections presented in Fig.~\ref{fig:msigma} they all overlap except for the 0 observed events bound in the left panel, and also in the right panel except at the bottom of the plot.

Fig.~\ref{fig:msigma} shows that, within the SHM both for ADM and MDM, the allowed regions do overlap, as just mentioned, but the regions are entirely rejected by the combined 90\% CL limits of LUX and CDMSlite, or the 90\% CL SuperCDMS limit by itself.

\section{Halo-independent analysis}\label{sec:halo-indep}

In this section we proceed as in Ref.~\cite{DelNobile:2013cva}. Changing the order of the $\bfv$ and $\ER$ integrations in Eq.~\eqref{eq:R} we can write 
\beq
\label{eq:Rred}
R_{[\Ed_1, \Ed_2]}(t) =  \int_0^\infty \ud^3 v \, \frac{\tilde{f}(\bfv, t)}{v} \, \eH_{[\Ed_1, \Ed_2]}(v) \ ,
\eeq
where we define
\beq
\label{eq:ftilde}
\tilde{f}(\bfv, t) \equiv \frac{\rho \sigma_{\rm ref}}{\mDM} \, f(\bfv, t)
\eeq
and
\begin{gather}
\label{eq:HT}
\eH_{[\Ed_1, \Ed_2]}(v) \equiv
\sum_T \frac{C_T}{m_T} \int_0^{\ER^{\rm max}(v)} \ud \ER \, \frac{v^2}{\sigma_{\rm ref}} \frac{\ud \sigma_T}{\ud \ER}(\ER, v) \int_{\Ed_1}^{\Ed_2} \ud\Ed \, \epsilon(\Ed) G_T(\ER, \Ed)
\\
= \sum_T \frac{C_T}{m_T} \frac{4 \mu_T^2}{m_T} \int_0^v \ud \vmin \, \vmin \frac{v^2}{\sigma_{\rm ref}} \frac{\ud \sigma_T}{\ud \ER}(\ER(\vmin), v) \int_{\Ed_1}^{\Ed_2} \ud\Ed \, \epsilon(\Ed) G_T(\ER(\vmin), \Ed) \ .
\nonumber
\end{gather}
Here $\ER^{\rm max}(v) \equiv 2 \mu_T^2 v^2 / m_T$ is the maximum recoil energy a WIMP of speed $v$ can impart in an elastic collision to a target nucleus initially at rest. We divided the cross section by $\sigma_{\rm ref}$, which contains the unknown coupling constants, to factor it into $\tilde{f}(\bfv, t)$, together with $\rho$. We also factored out from $\eH$ a $1/v^2$ factor in order to write the velocity integral in 
Eq.~\eqref{eq:Rred} in the usual form $\tilde{f}(\bfv, t) / v$ of the rate of simple contact interactions. Finally, in the second line we merely changed integration variable from $\ER$ to $\vmin$ using Eq.~\eqref{eq:vmin}.

Introducing the speed distribution
\begin{align}
\widetilde{F}(v, t) \equiv v^2 \int \ud\Omega_v \, \tilde{f}(\bfv, t)\ ,
\end{align}
we can rewrite Eq.~\eqref{eq:Rred} as
\begin{align}
R_{[\Ed_1, \Ed_2]}(t) & =  \int_0^\infty \ud v \, \frac{\widetilde{F}(v, t)}{v} \, \eH_{[\Ed_1, \Ed_2]}(v) \ .
\label{eq:REEbis}
\end{align}
We now define the function $\tilde{\eta}(v, t)$ by
\begin{align}
\label{eta-derivative}
\frac{\widetilde{F}(v, t)}{v}  = - \left. \frac{ \partial \tilde{\eta}(\vmin, t) }{\partial \vmin} \right|_{\vmin = v} \ ,
& & \text{\ie} & &
\tilde{\eta}(\vmin, t) = \int_{\vmin}^\infty \ud v \, \frac{\widetilde{F}(v, t)}{v}\ .
\end{align}
$\tilde{\eta}(\vmin, t)$ goes to zero in the limit of $\vmin$ going to infinity. Using Eq.~\eqref{eta-derivative} in Eq.~\eqref{eq:REEbis} we get therefore
\begin{align}
\label{eq:R3}
R_{[\Ed_1, \Ed_2]}(t) & = - \int_0^\infty \ud v \, \frac{ \partial \tilde{\eta}(v, t) }{\partial v}  \, \eH_{[\Ed_1, \Ed_2]}(v) = \int_0^\infty \ud\vmin \, \tilde{\eta}(\vmin, t) \,  \eR_{[\Ed_1, \Ed_2]}(\vmin) \ ,
\end{align}
where in the last equality we integrated by parts and defined the response function
\begin{align}
\label{eq:RT}
\eR_{[\Ed_1, \Ed_2]}(\vmin) \equiv \left. \frac{\partial \eH_{[\Ed_1, \Ed_2]}(v)}{\partial v} \right|_{v = \vmin} \ .
\end{align}
Notice that the boundary term in the integration by parts is zero because $\eH_{[\Ed_1, \Ed_2]}(0) = 0$ by definition in Eq.~\eqref{eq:HT} (since $\ER^{\rm max}(0) = 0$). More explicitly, the response function has the form
\begin{multline}
\eR_{[\Ed_1, \Ed_2]}(\vmin) =
\sum_T \frac{C_T}{m_T} \frac{4 \mu_T^2}{m_T} \left\{
\frac{\vmin^3}{\sigma_{\rm ref}} \frac{\ud \sigma_T}{\ud \ER}(\ER(\vmin), \vmin) \int_{\Ed_1}^{\Ed_2} \ud\Ed \, \epsilon(\Ed) G_T(\ER(\vmin), \Ed) \right. +
\\
\left. \int_0^{\vmin} \ud w \, w \, \frac{\ud}{\ud \vmin} \left[ \frac{\vmin^2}{\sigma_{\rm ref}} \frac{\ud \sigma_T}{\ud \ER}(\ER(w), \vmin) \right] \int_{\Ed_1}^{\Ed_2} \ud\Ed \, \epsilon(\Ed) G_T(\ER(w), \Ed)
\right\} \ .
\end{multline}

The velocity integral $\tilde{\eta}(\vmin, t)$ has an annual modulation due to the revolution of Earth around the Sun typically well approximated by the first terms of a harmonic series,
\beq\label{etat}
\tilde{\eta}(\vmin, t) \simeq \tilde{\eta}^0(\vmin) + \tilde{\eta}^1(\vmin) \cos\!\left[ \omega (t - t_0) \right] \ ,
\eeq
where $t_0$ is the time of maximum signal and $\omega = 2 \pi/$yr. The unmodulated and modulated components $\tilde{\eta}^0$ and $\tilde{\eta}^1$ enter respectively in the definition of the unmodulated and modulated parts of the rate,
\beq\label{Rt}
R_{[\Ed_1, \Ed_2]}(t) = R^0_{[\Ed_1, \Ed_2]} + R^1_{[\Ed_1, \Ed_2]} \cos\!\left[ \omega (t - t_0) \right] \ .
\eeq
Once the WIMP mass is fixed, the functions $\tilde{\eta}^0(\vmin)$ and $\tilde{\eta}^1(\vmin)$ are detector-independent quantities that must be common to all (non-directional) direct DM experiments. Therefore we can map the rate measurements and bounds of different experiments into measurements of and bounds on $\tilde{\eta}^0(\vmin)$ and $\tilde{\eta}^1(\vmin)$ as functions of $\vmin$.

Following Ref.~\cite{DelNobile:2013cva} we map rate measurements $\hat{R}^i_{[\Ed_1, \Ed_2]}$ into measurements of $\tilde{\eta}^i(\vmin)$ ($i=0$, $1$) as follows. From Eq.~\eqref{eq:R3} we get averages of $\tilde{\eta}^i(\vmin)$ with a weight $\eR_{[\Ed_1, \Ed_2]}(\vmin)$
\beq
\overline{\tilde{\eta}^i_{[\Ed_1, \Ed_2]}(\vmin)}
\equiv
\frac{\hat{R}^i_{[\Ed_1, \Ed_2]}}{\int_0^\infty \ud\vmin \, \eR_{[\Ed_1, \Ed_2]}(\vmin)}
=
\frac{\hat{R}^i_{[\Ed_1, \Ed_2]}}{\eH_{[\Ed_1, \Ed_2]}(\infty)} \ ,
\eeq
where $\overline{\tilde{\eta}^i_{[\Ed_1, \Ed_2]}(\vmin)}$ indicates the average value of $\tilde{\eta}^i(\vmin)$ in the $\vmin$ range corresponding to the width of the response function $\eR_{[\Ed_1, \Ed_2]}(\vmin)$. For interactions whose differential cross section goes as $v^{-2}$, e.g. for the simplest examples of contact interaction, the only dependence of the function $\eH_{[\Ed_1, \Ed_2]}(v)$ on $v$ is through the upper limit of the $\ER$ integral, $\ER^{\rm max}(v)$ (see Eq.~\eqref{eq:HT}). Thus the response function $\eR_{[\Ed_1, \Ed_2]}(\vmin)$ is proportional to $\ud\ER^{\rm max} / \ud v$, which grows linearly with $v$, times a function that vanishes rapidly for large enough $\ER^{\rm max}(v)$ due to the detector resolution function $G_T(\ER, \Ed)$ being localized. For this reason, in this case $\eR$ has a finite width and its integral is finite.

\begin{figure}[t]
\centering
\includegraphics[width=0.49\textwidth]{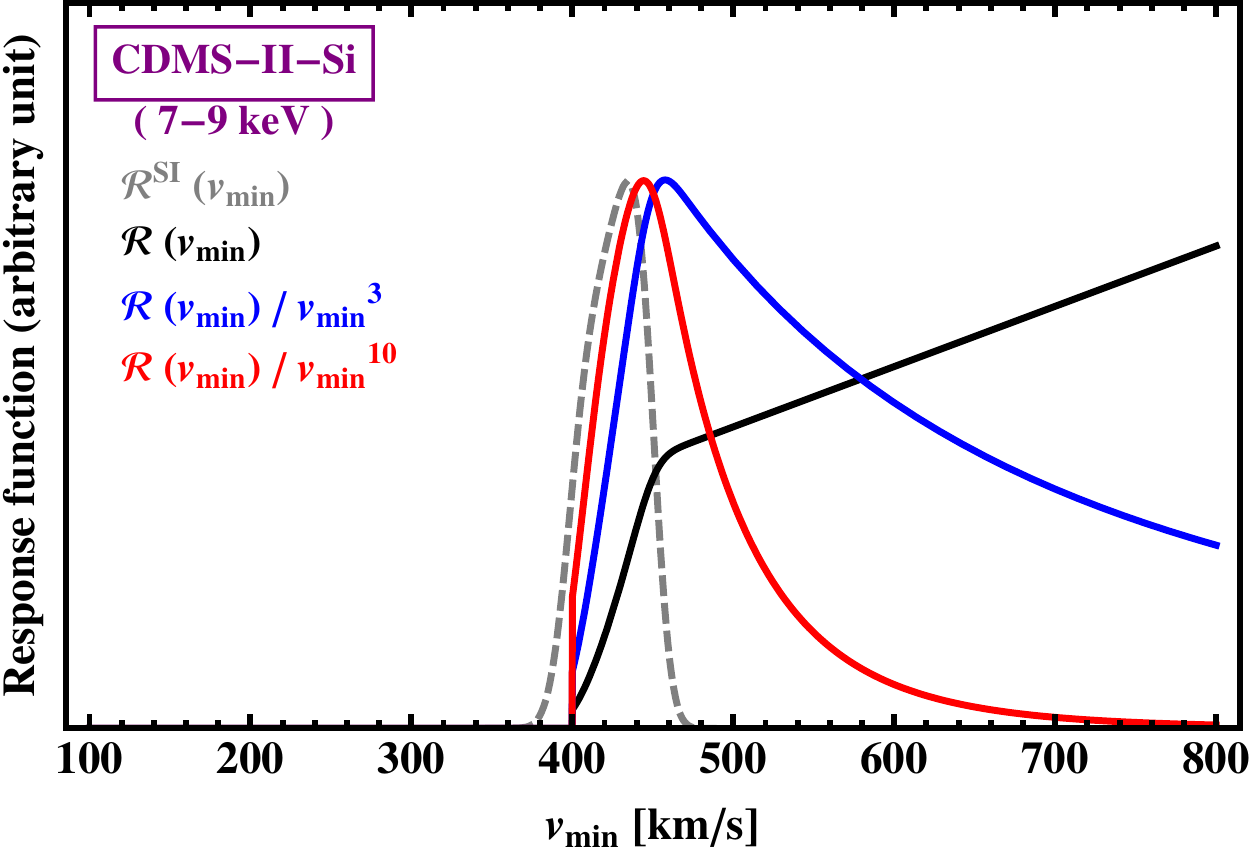}
\includegraphics[width=0.49\textwidth]{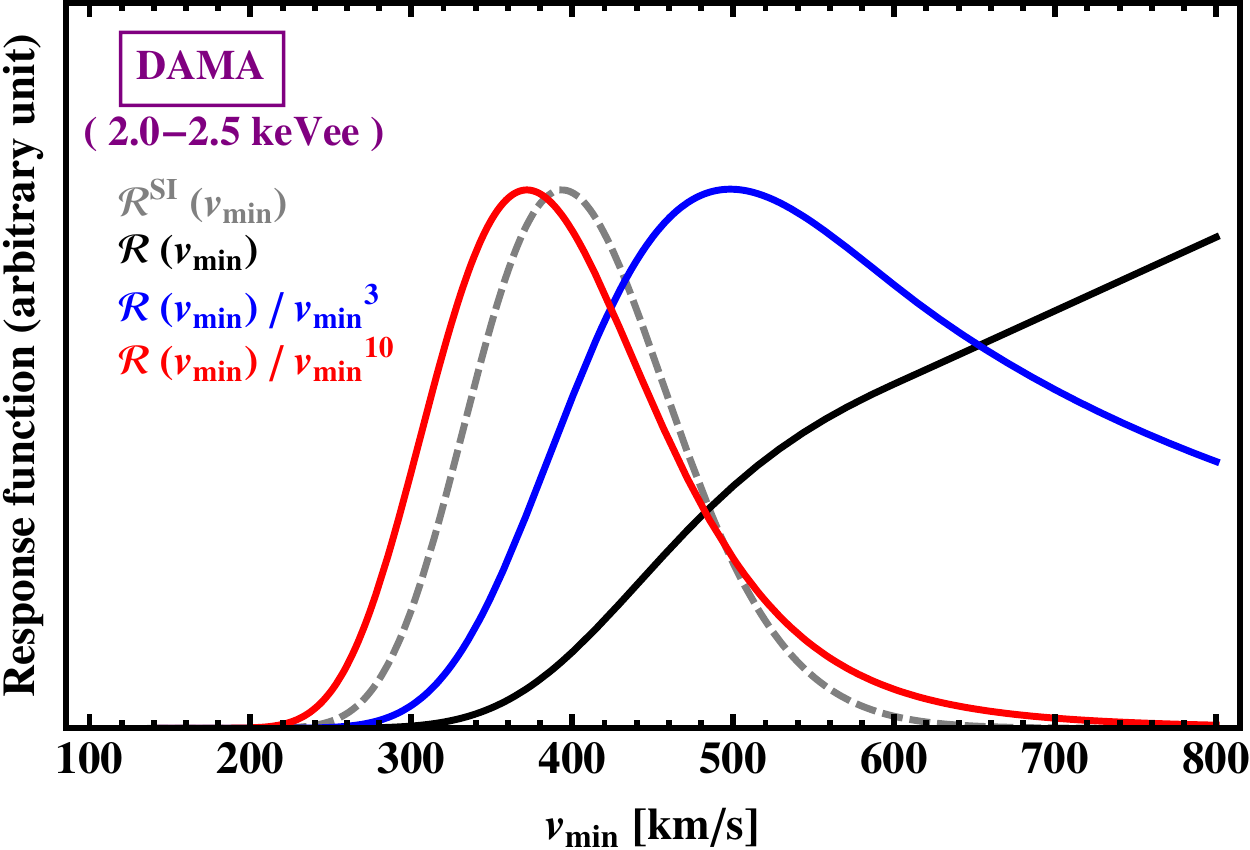}
\includegraphics[width=0.49\textwidth]{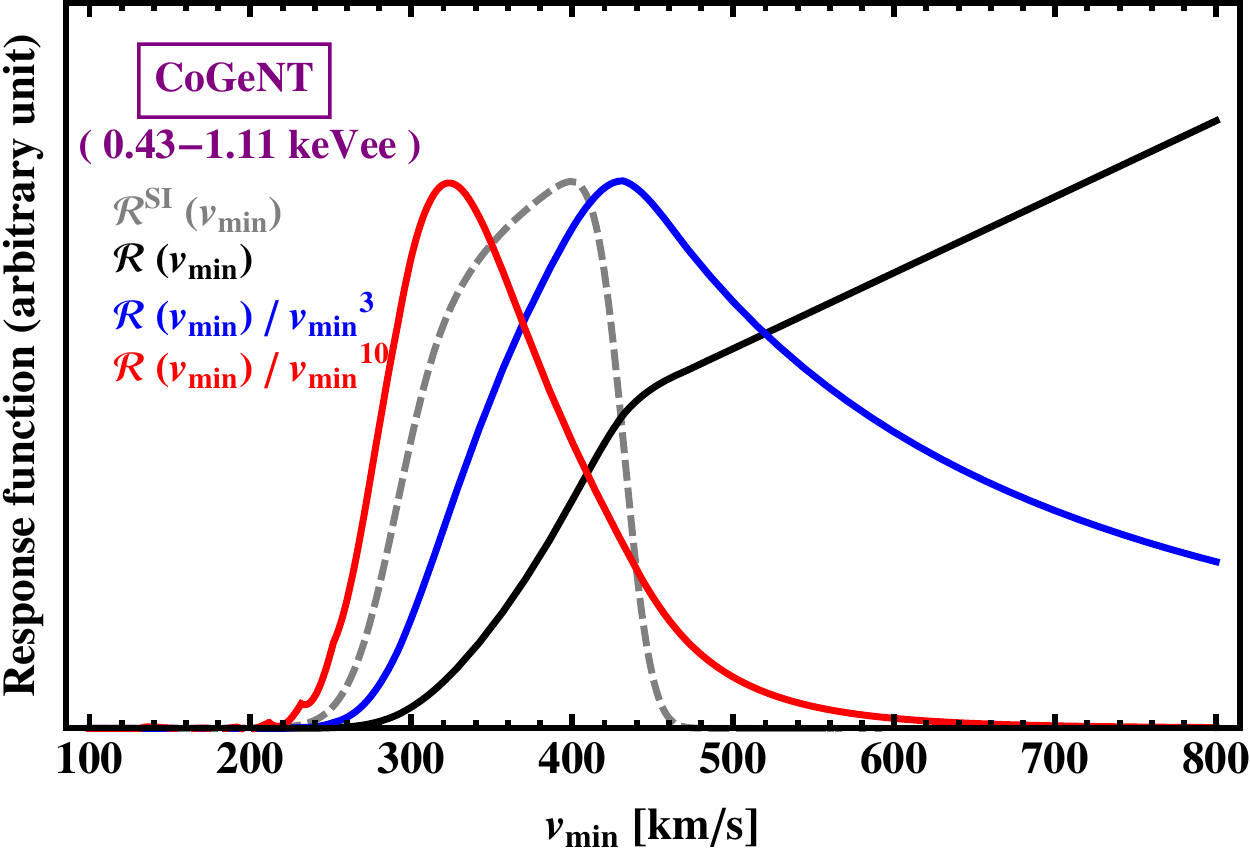}
\includegraphics[width=0.49\textwidth]{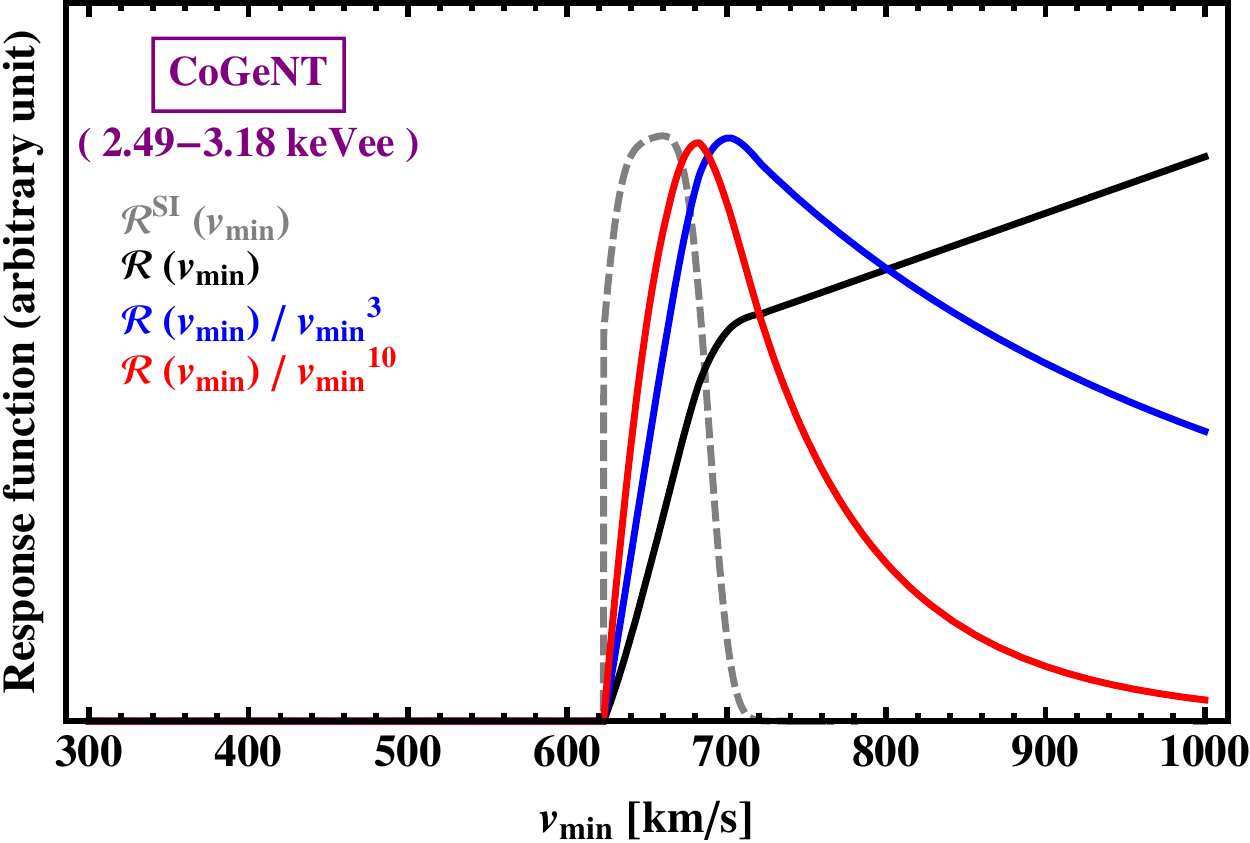}
\caption{\label{fig:R}
Modified response functions $\vmin^{-r} \eR_{[\Ed_1, \Ed_2]}(\vmin)$ for ADM with $m=10 \text{ GeV}/c^2$ with arbitrary normalization, for several detected energy intervals and detectors. The colored continuous curves are for various values of $r$. The gray dashed curve corresponds the usual spin-independent contact-interaction and has been included here for comparison. }
\end{figure}

When $\ud\sigma_T / \ud\ER$ depends on powers of $v$ higher than $-2$ as is the case of ADM and MDM where a constant term $v^0$ is present, the integral of the response function $\eH$ is infinite. In Ref.~\cite{DelNobile:2013cva} this problem was circumvented by \virg{regularizing} the average, \ie by considering the average of $\vmin^r \tilde{\eta}^i(\vmin)$ for some suitable positive power $r$ such that $\int_0^\infty \ud\vmin \, \vmin^{-r} \eR_{[\Ed_1, \Ed_2]}(\vmin)$ is finite. Thus, we compare
\beq\label{newaverage}
\overline{\vmin^r \tilde{\eta}^i_{[\Ed_1, \Ed_2]}(\vmin)}
\equiv
\frac{\hat{R}^i_{[\Ed_1, \Ed_2]}}{\int_0^\infty \ud\vmin \, \vmin^{-r} \eR_{[\Ed_1, \Ed_2]}(\vmin)}
=
\frac{\hat{R}^i_{[\Ed_1, \Ed_2]}}{r \int_0^\infty \ud v \, v^{-r-1} \eH_{[\Ed_1, \Ed_2]}(v)}
\eeq
for different experiments. In the last equality we integrated by parts and used the fact that the boundary term $\left[ v^{-r} \, \eH_{[\Ed_1, \Ed_2]}(v) \right]^\infty_0$ vanishes for a good choice of $r$. In this way one obtains a \virg{modified response function} $\vmin^{-r} \eR_{[\Ed_1, \Ed_2]}(\vmin)$ with finite width. $r$ should be large enough so that the $\vmin$ ranges of different energy intervals do not overlap too much, but small enough not to give a large weight to the low velocity tail of the response function, which depends on the low energy tail of the resolution function $G_T(\ER, \Ed)$ that is never well known. In Fig.~\ref{fig:R} we show for ADM some modified response functions $\vmin^{-r} \eR_{[\Ed_1, \Ed_2]}(\vmin)$ for various values of $r$. While the minimum $r$ required to regularize the integral for ADM is $r=2$, we see that the modified response function is very wide for $r$ values up to about $10$ (the same as for MDM, see Ref.~\cite{DelNobile:2013cva}). We use thus $r=10$ in the figures (same choice as for MDM in Ref.~\cite{DelNobile:2013cva}).

\begin{figure}[t]
\centering
\includegraphics[width=0.49\textwidth]{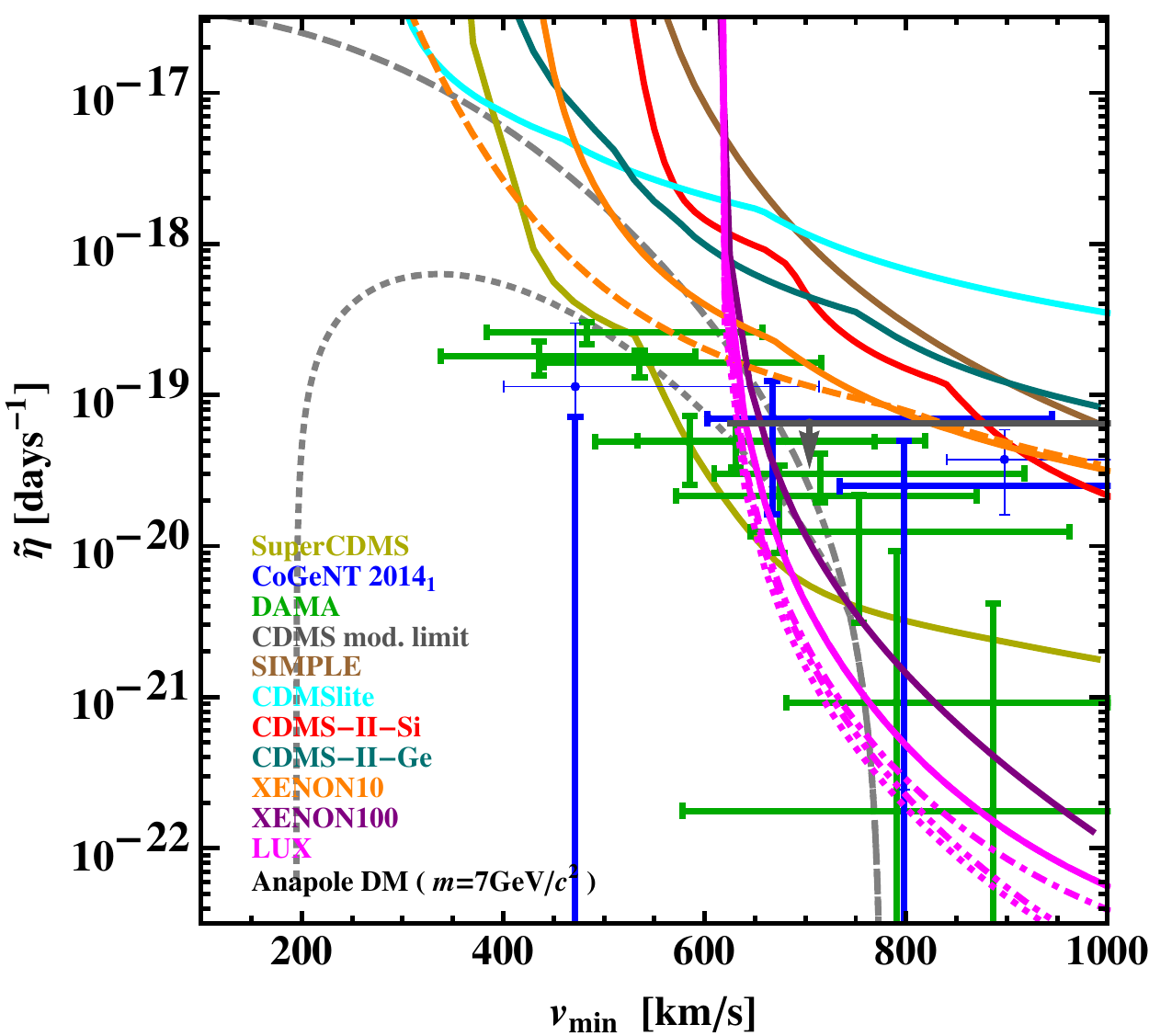}
\includegraphics[width=0.49\textwidth]{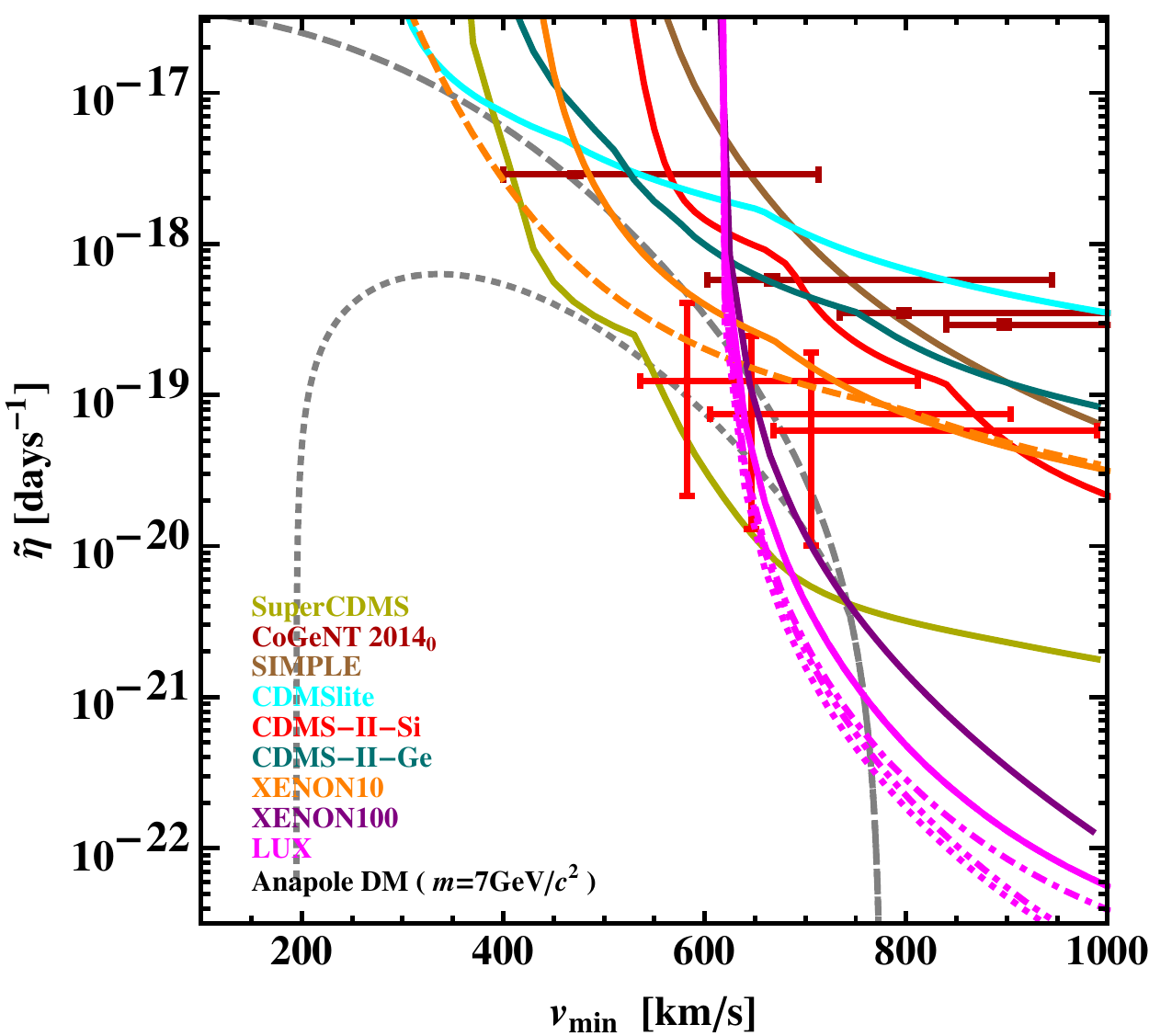}
\\
\includegraphics[width=0.49\textwidth]{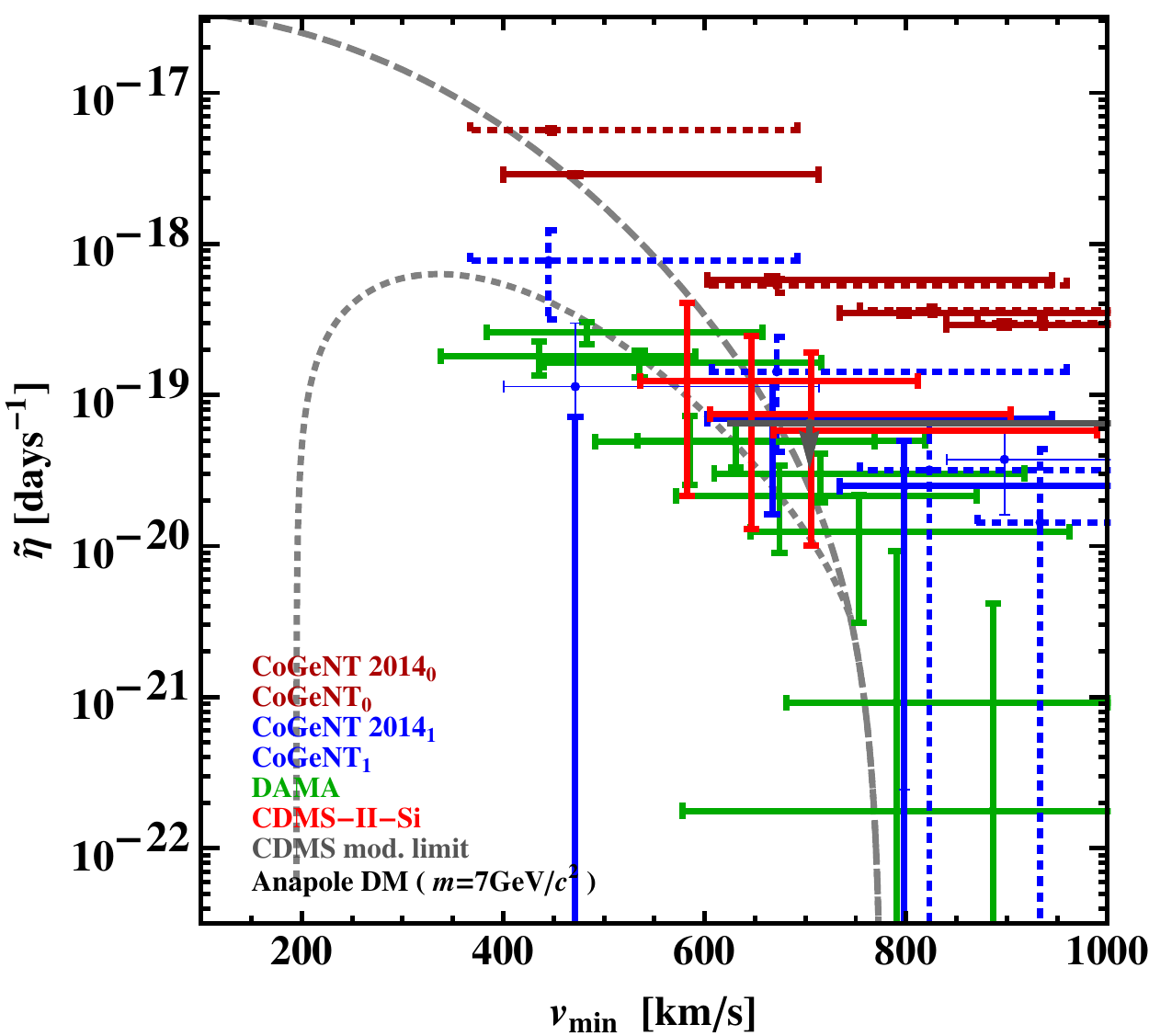}
\includegraphics[width=0.49\textwidth]{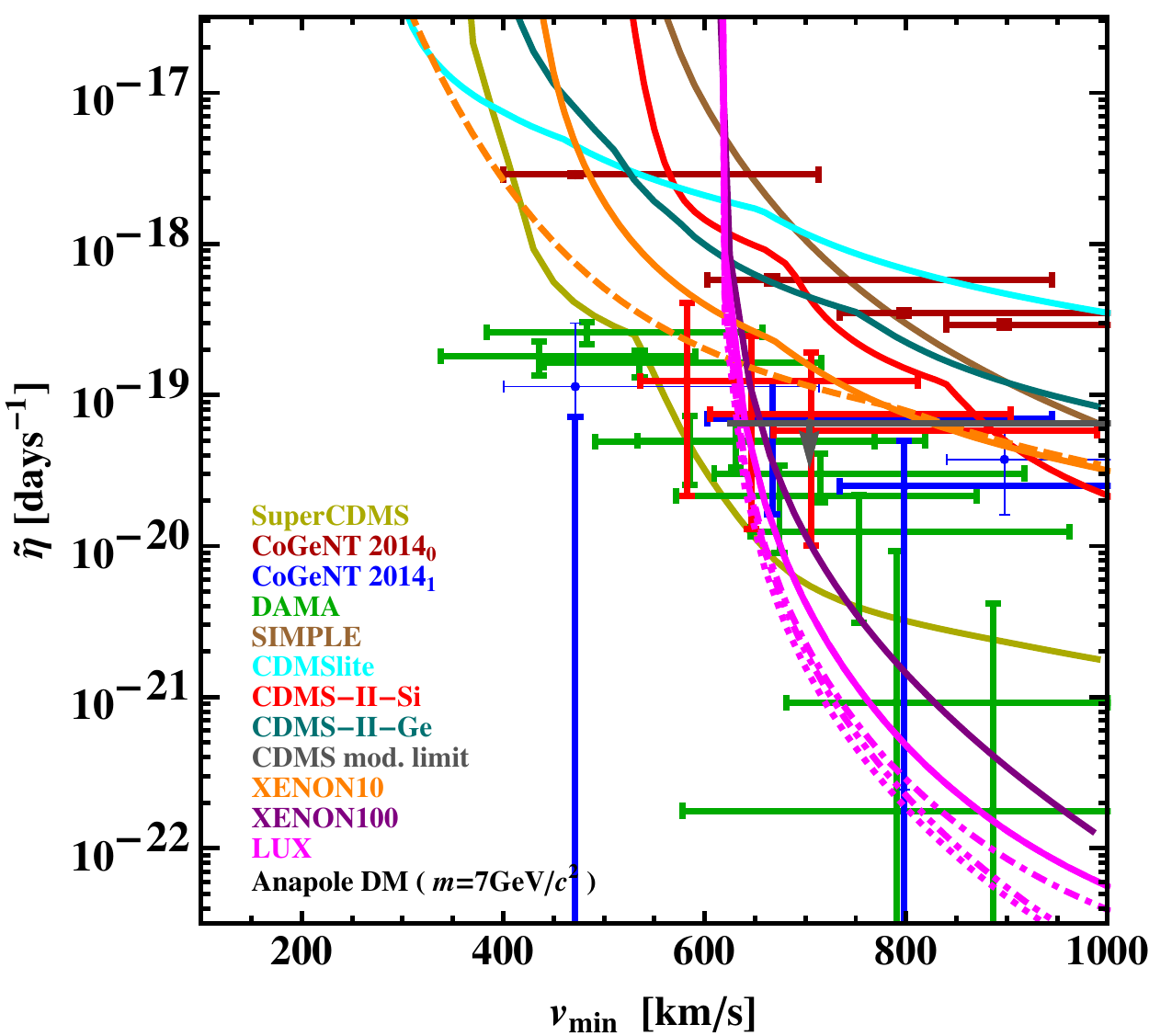}
\caption{\label{fig:eta7}
Measurements of and upper bounds on 
$\vmin^{-10} \overline{\vmin^{10} \tilde\eta^0(\vmin)} c^2$ and $\vmin^{-10} \overline{\vmin^{10} \tilde\eta^1(\vmin)} c^2$ for ADM with $m = 7$ GeV/$c^2$. The sodium quenching factor in DAMA is assumed to be $Q_{\rm Na} = 0.30$. The different lines for XENON10 and LUX are as explained in the text. For the first and fourth CoGeNT 2014 modulation data points we show the modulus of the negative part of the cross with a thin blue line. To avoid excessive cluttering we separate the material in four panels. {\it Top left:} bounds on $\vmin^{-10} \overline{\vmin^{10} \tilde\eta^0(\vmin)} c^2$ (lines) and measurements of $\vmin^{-10} \overline{\vmin^{10} \tilde\eta^1(\vmin)} c^2$ (crosses). {\it Top right:} bounds on and measurements of $\vmin^{-10} \overline{\vmin^{10} \tilde\eta^0(\vmin)} c^2$. {\it Bottom left:} measurements of $\vmin^{-10} \overline{\vmin^{10} \tilde\eta^0(\vmin)} c^2$ and $\vmin^{-10} \overline{\vmin^{10} \tilde\eta^1(\vmin)} c^2$, and only the CDMS-Ge modulation limit. {\it Bottom right:} all the previous bounds and measurements shown together. In the bottom left panel we also show the CoGeNT 2011-2012 measurements (dashed crosses), for a comparison. In all the panels but the last one, the dashed gray lines show the SHM $\tilde{\eta}^0 c^2$ (upper line) and $\tilde{\eta}^1 c^2$ (lower line) for $\sigma_{\rm ref}^A = 3 \times 10^{-34}$ cm$^2$.}
\end{figure}

\begin{figure}[t]
\centering
\includegraphics[width=0.49\textwidth]{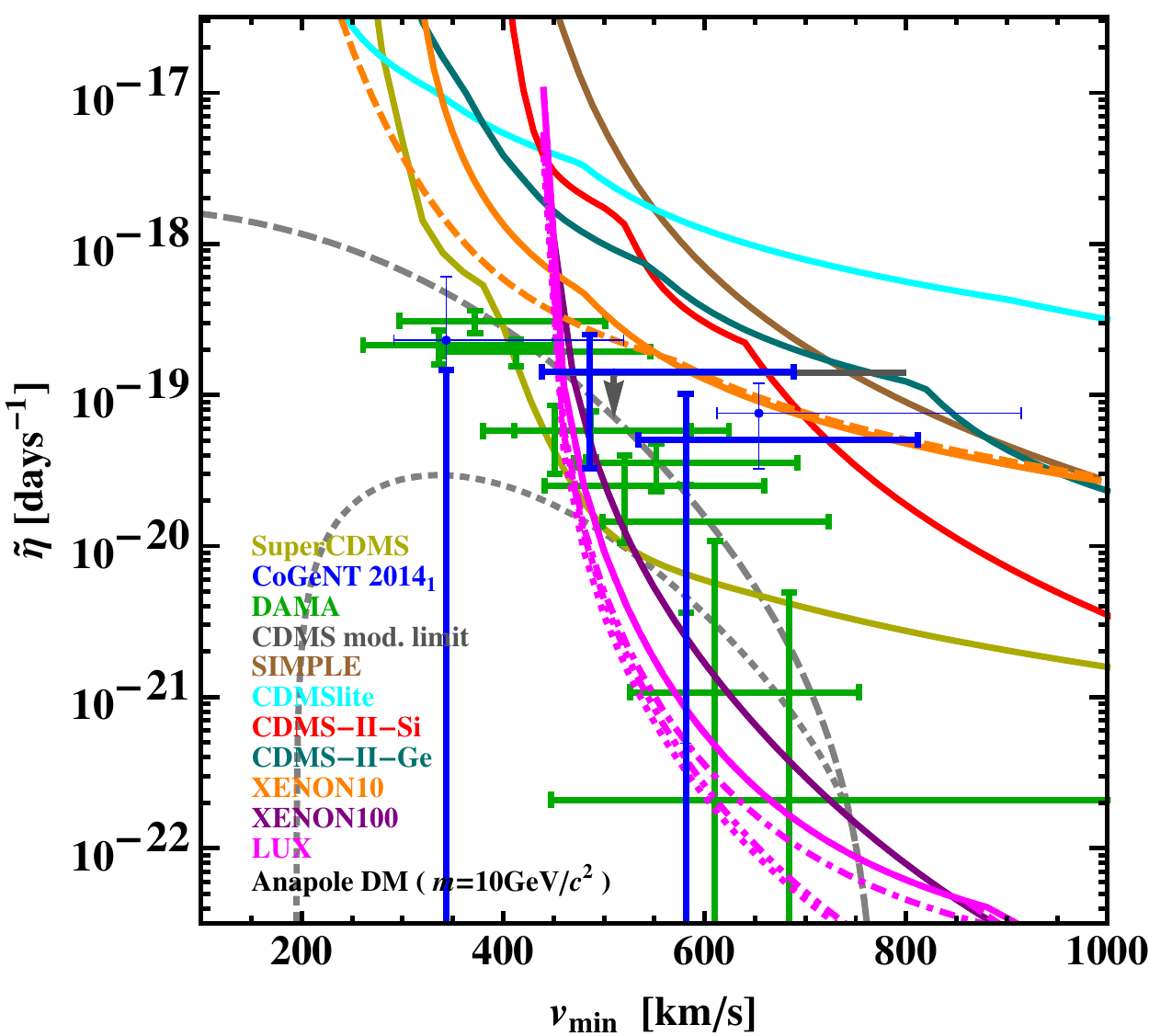}
\includegraphics[width=0.49\textwidth]{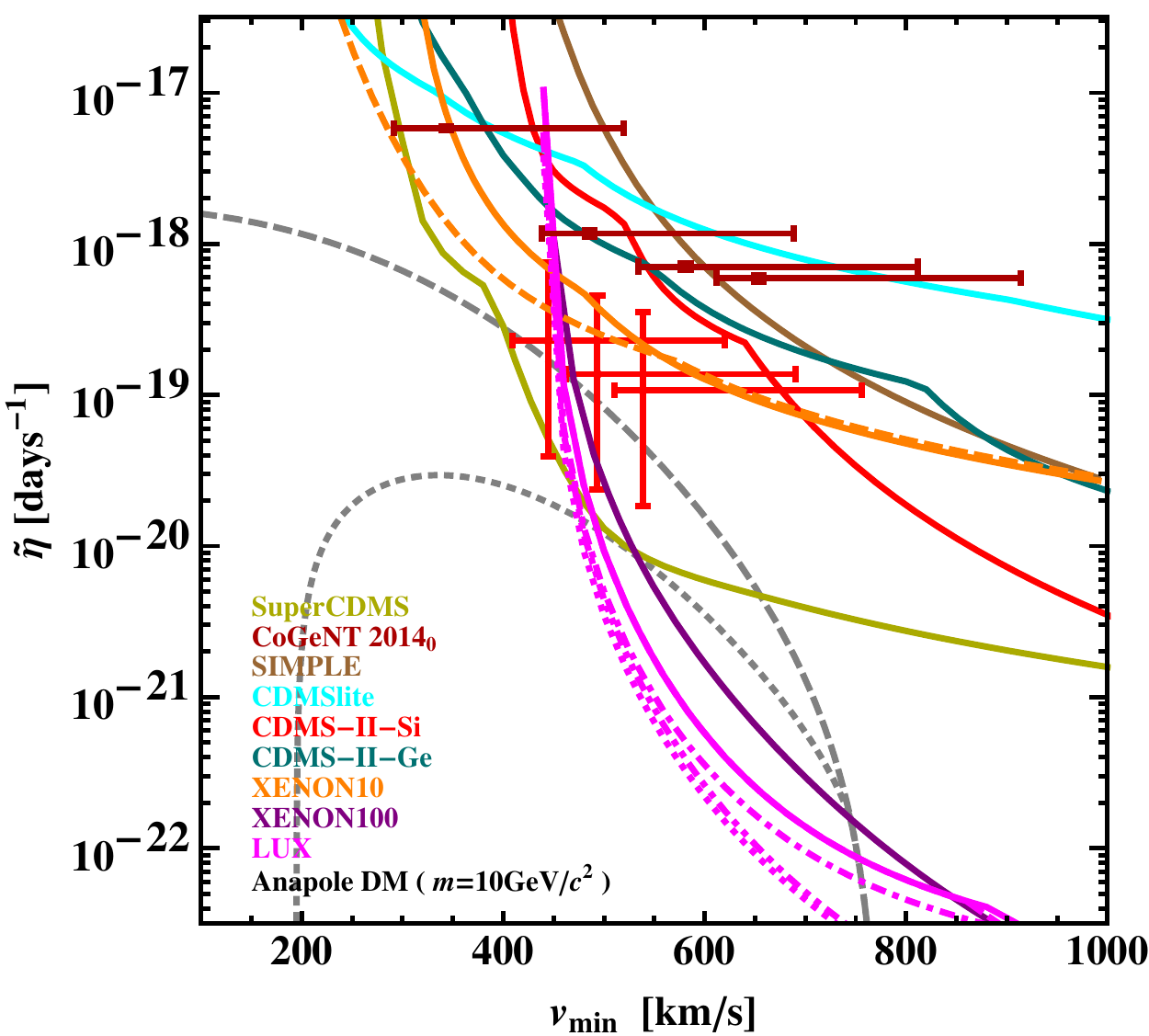}
\\
\includegraphics[width=0.49\textwidth]{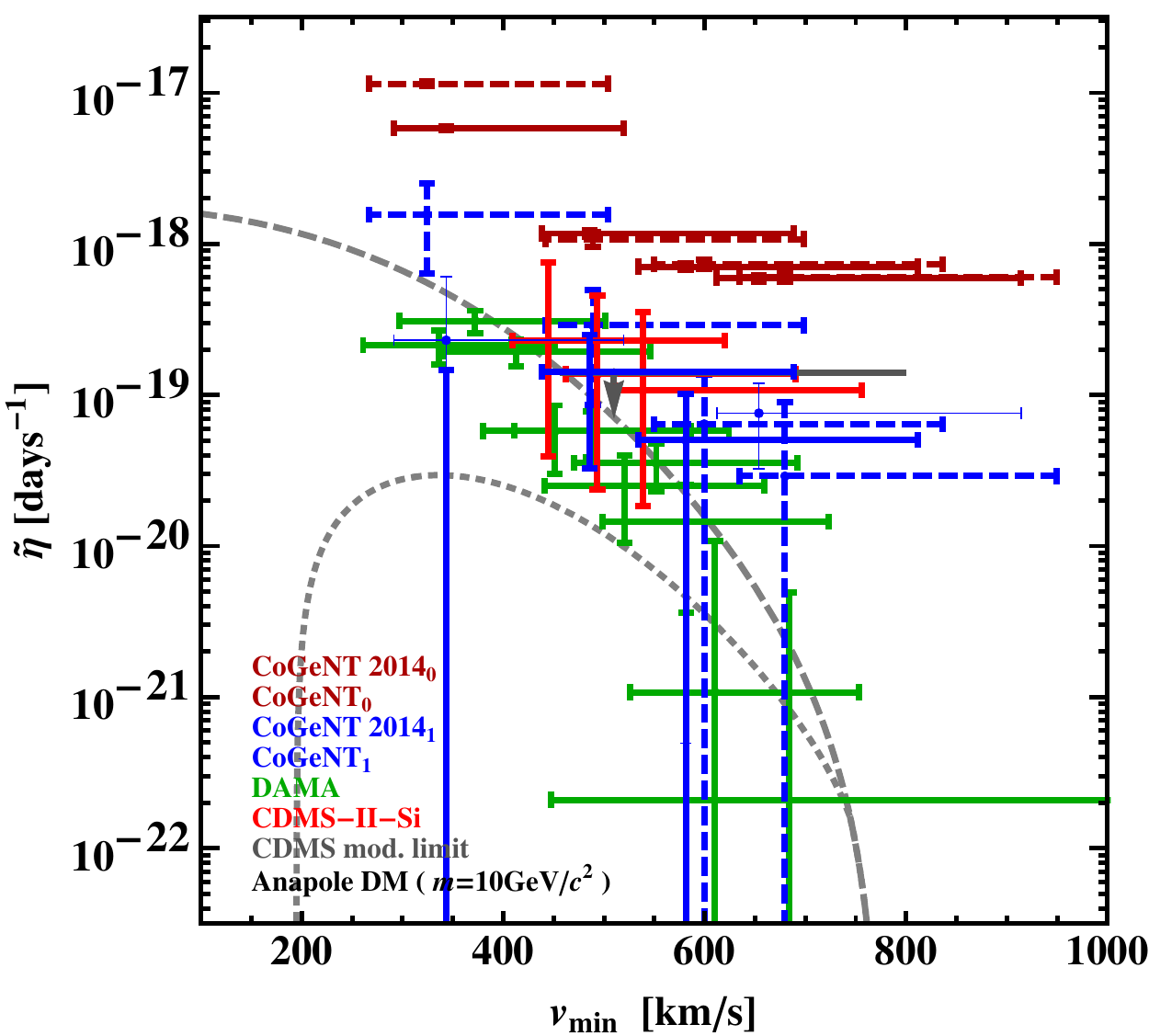}
\includegraphics[width=0.49\textwidth]{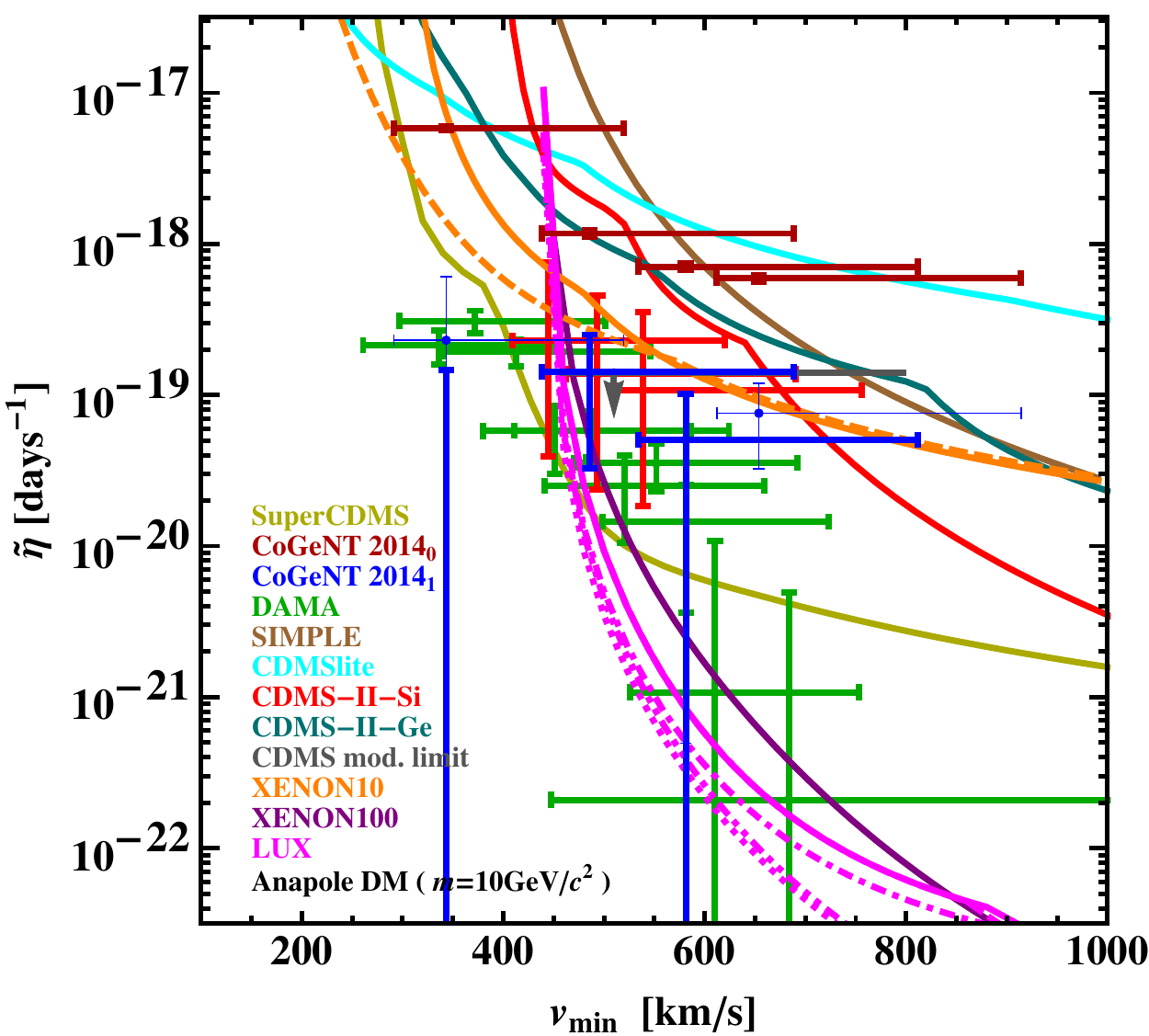}
\caption{\label{fig:eta10}
Same as in Fig.~\ref{fig:eta7}, but for $m = 10$ GeV/$c^2$. The dashed gray lines show $\tilde{\eta}^0 c^2$ (upper line) and $\tilde{\eta}^1 c^2$ (lower line) in the SHM for $\sigma_{\rm ref}^A = 2 \times 10^{-35}$ cm$^2$.}
\end{figure}

\begin{figure}[t]
\centering
\includegraphics[width=0.49\textwidth]{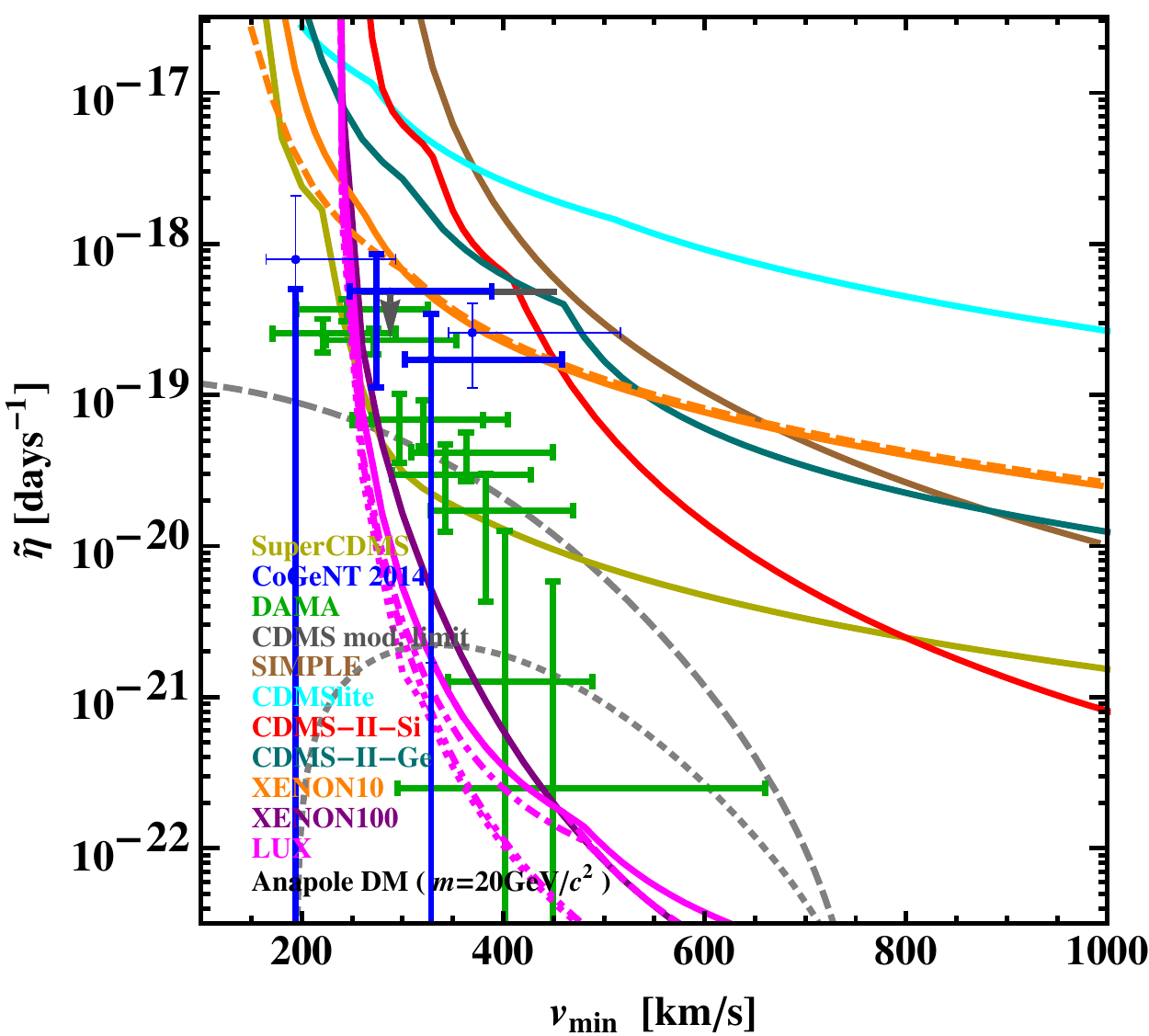}
\includegraphics[width=0.49\textwidth]{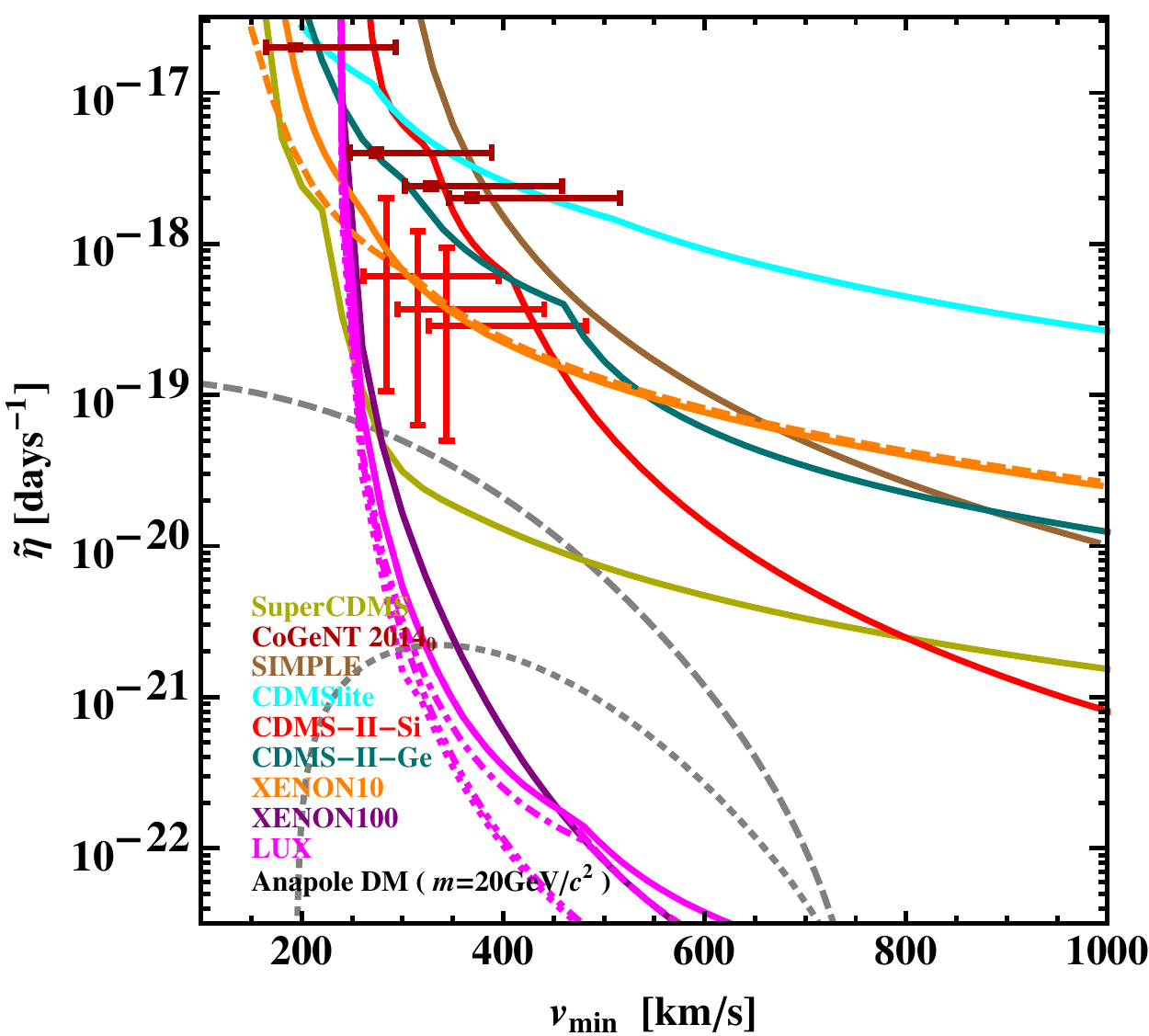}
\\
\includegraphics[width=0.49\textwidth]{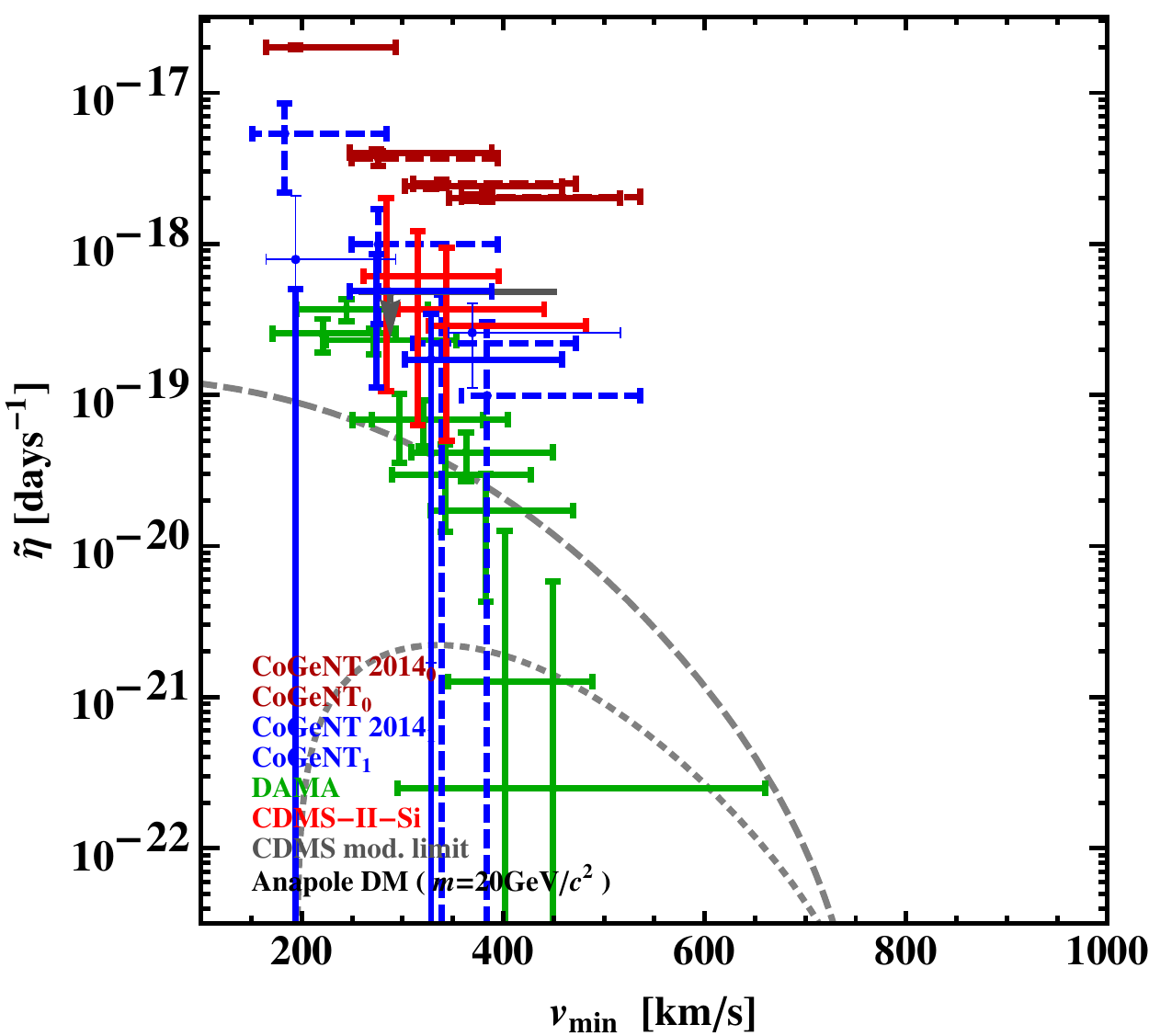}
\includegraphics[width=0.49\textwidth]{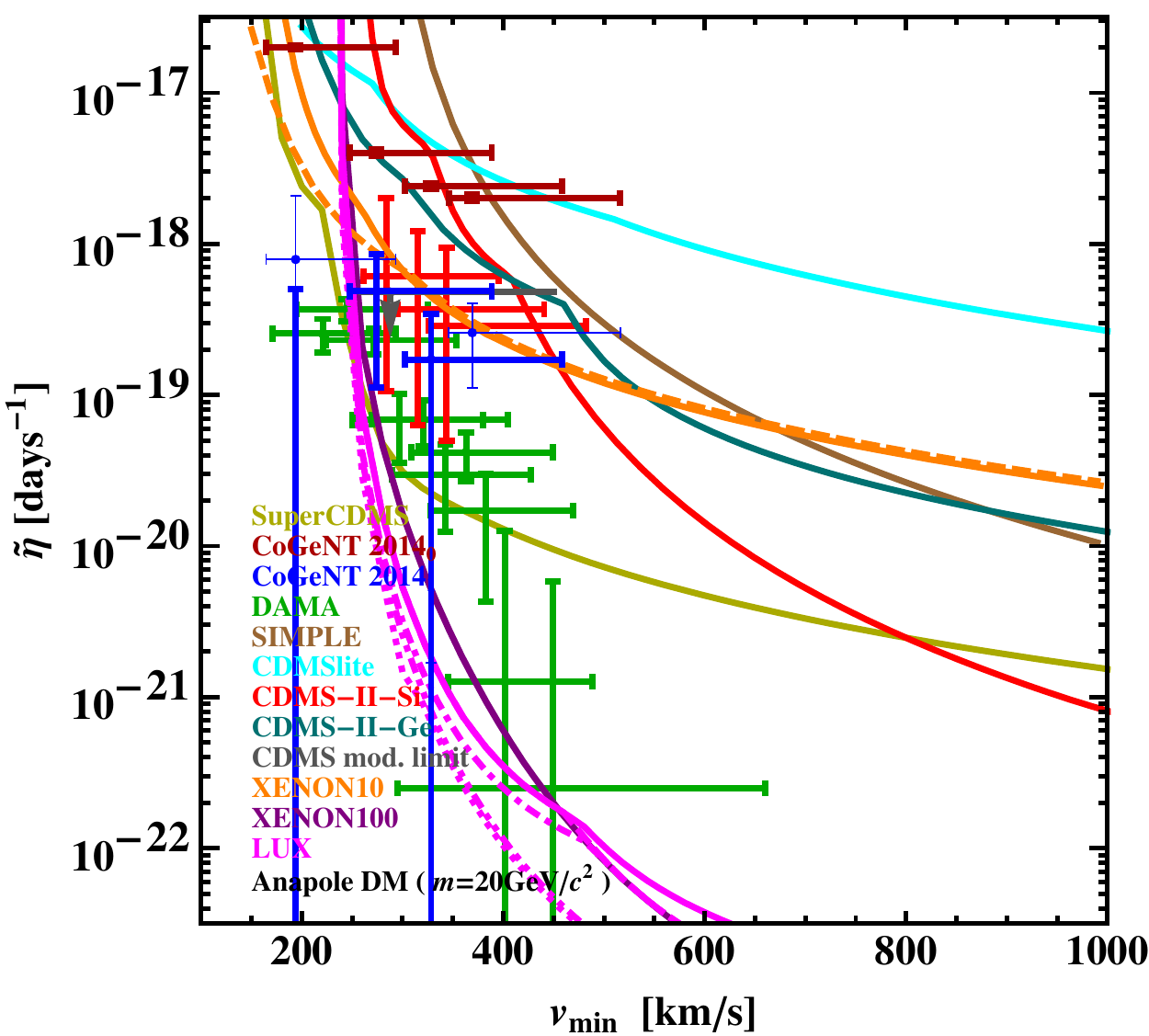}
\caption{\label{fig:eta20}
Same as in Fig.~\ref{fig:eta7}, but for $m = 20$ GeV/$c^2$. The dashed gray lines show $\tilde{\eta}^0 c^2$ (upper line) and $\tilde{\eta}^1 c^2$ (lower line) in the SHM for $\sigma_{\rm ref}^A = 3 \times 10^{-36}$ cm$^2$.}
\end{figure}

\begin{figure}[t]
\centering
\includegraphics[width=0.49\textwidth]{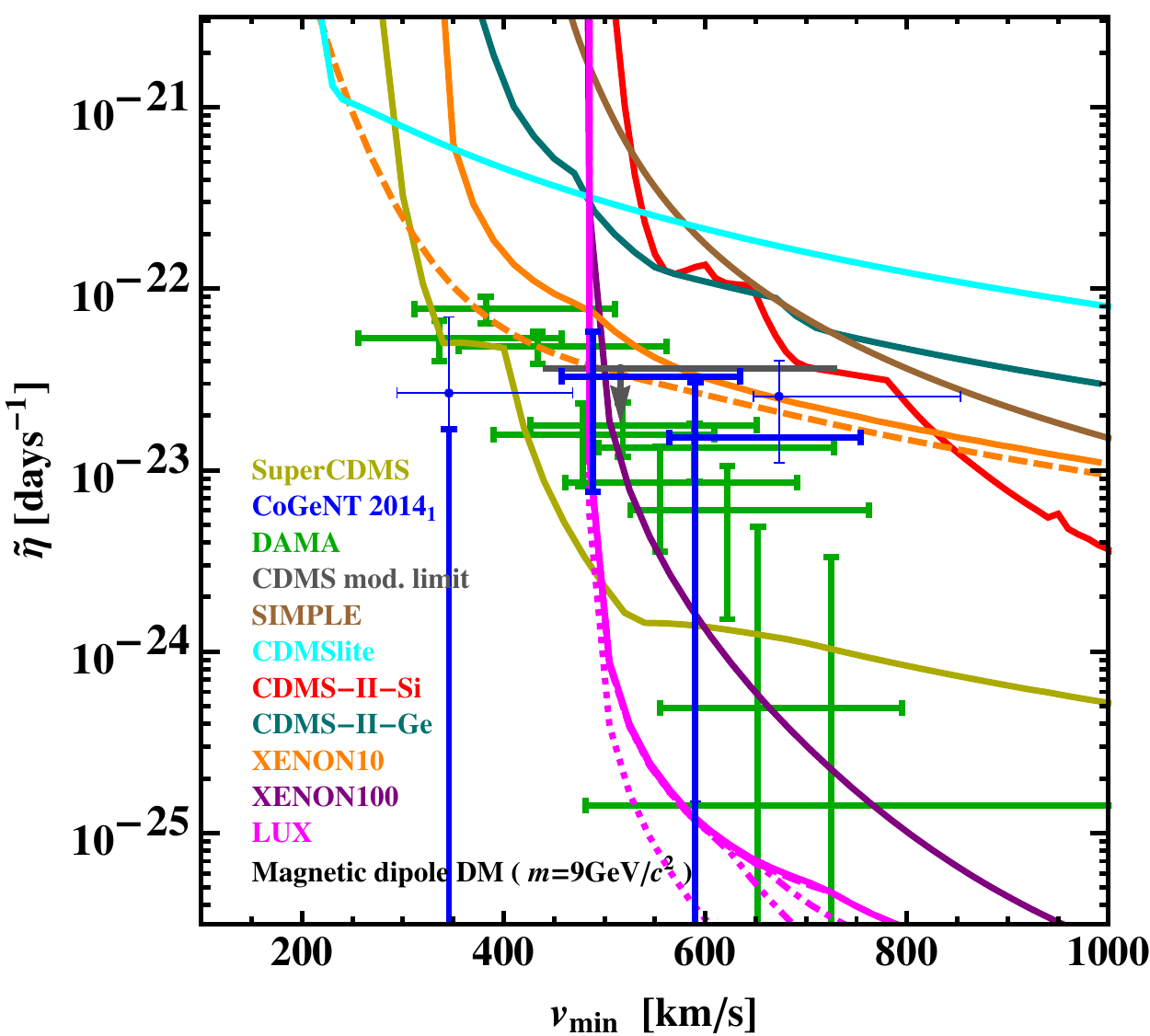}
\includegraphics[width=0.49\textwidth]{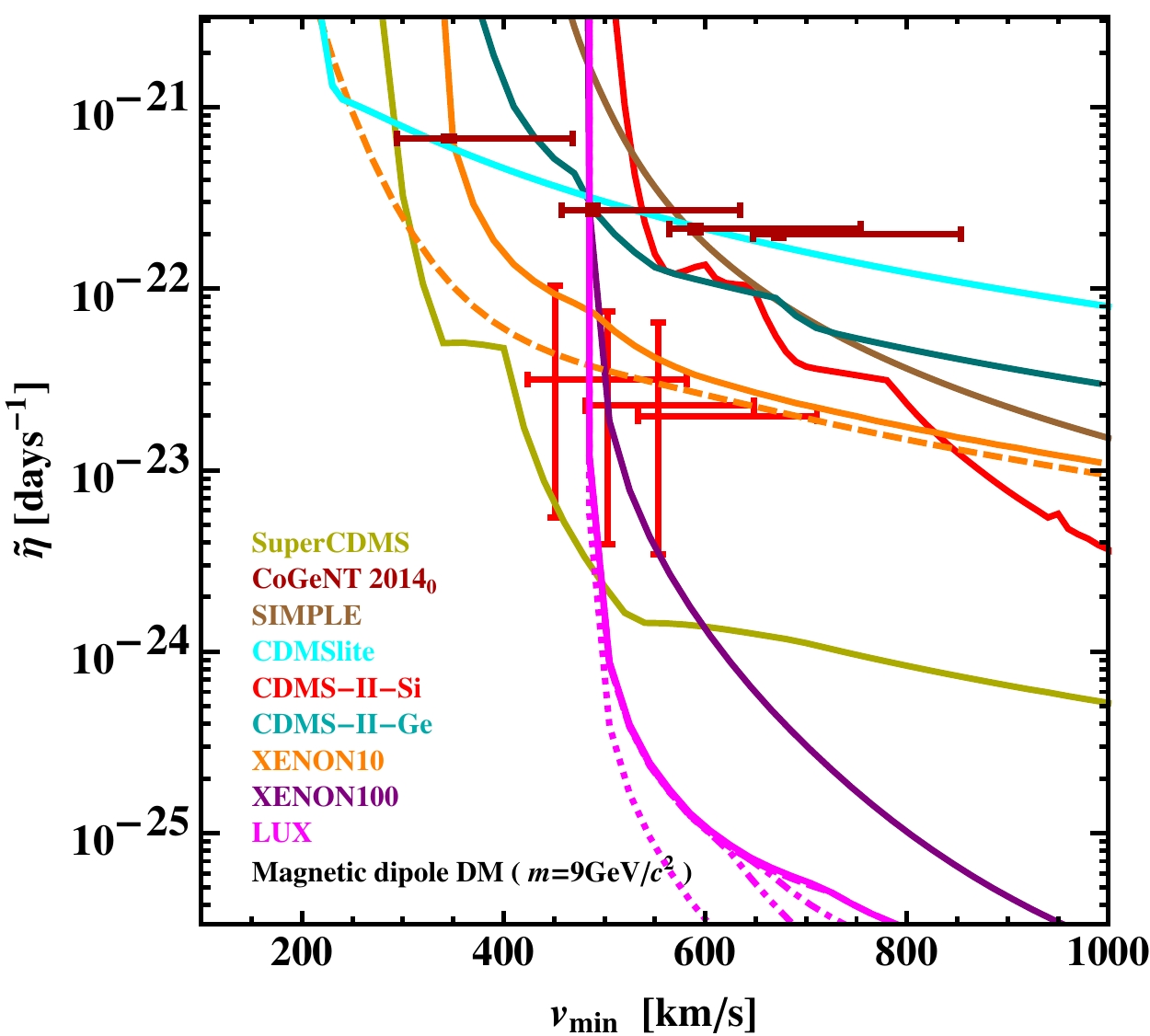}
\\
\includegraphics[width=0.49\textwidth]{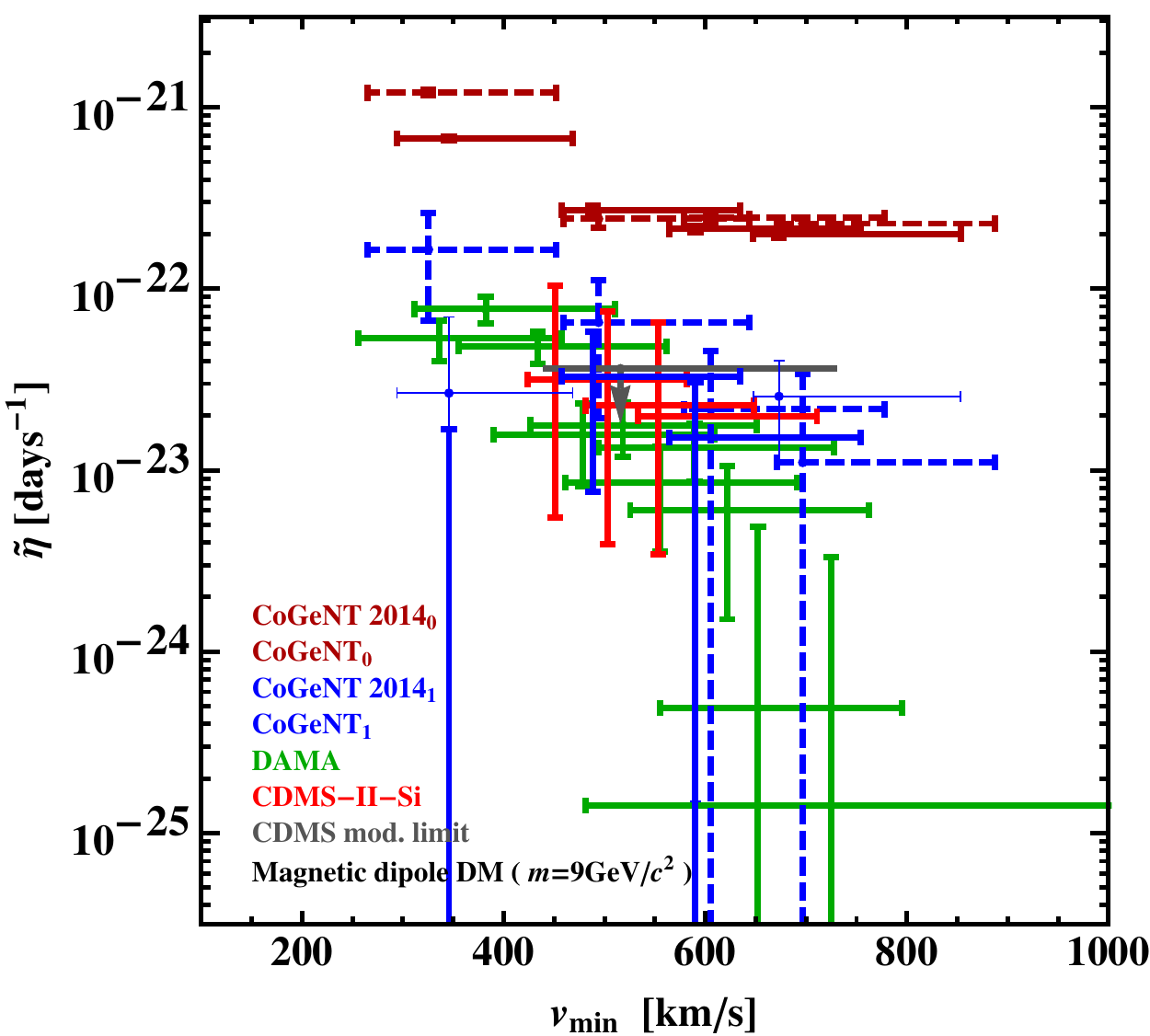}
\includegraphics[width=0.49\textwidth]{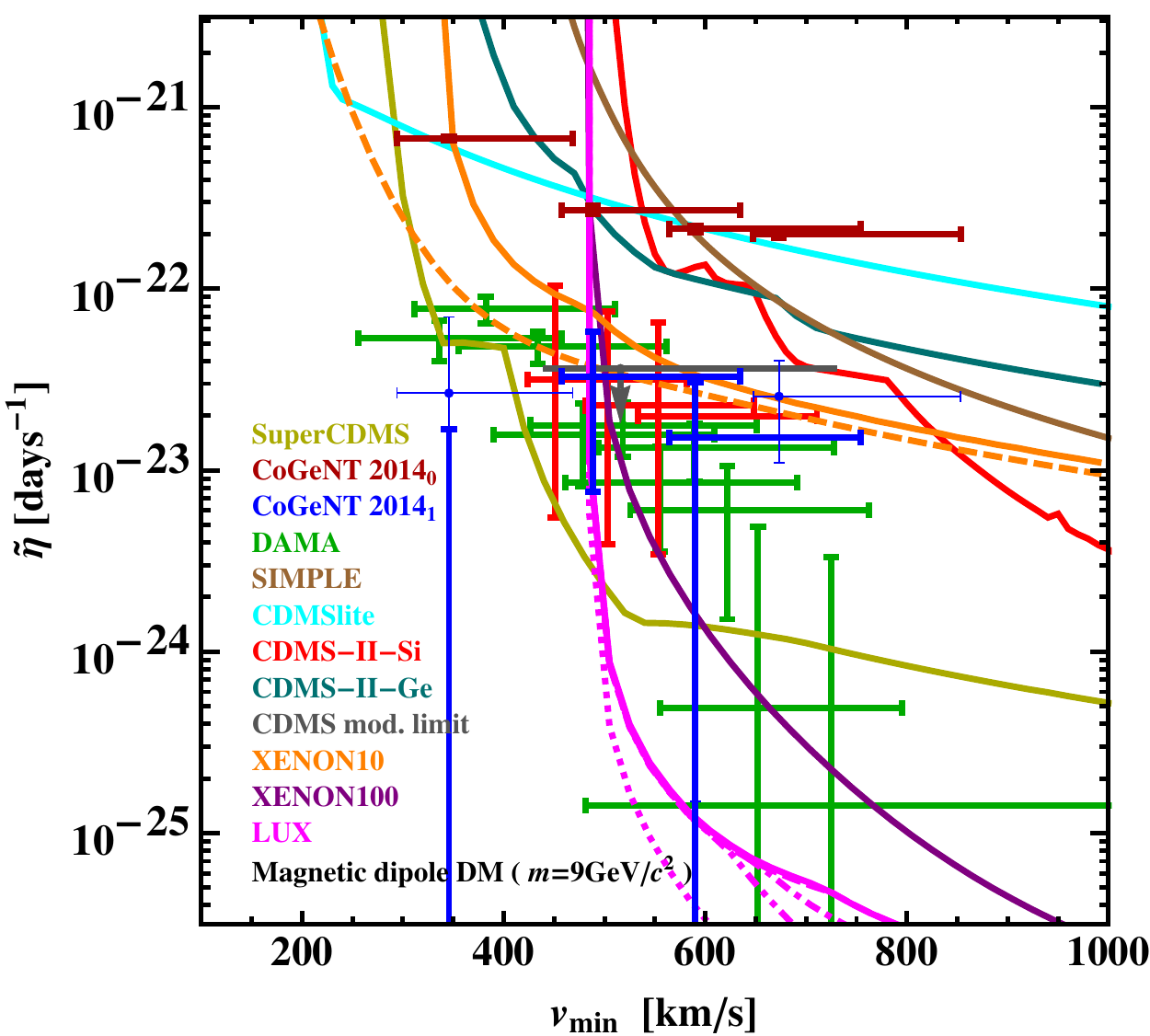}
\caption{\label{fig:eta9MDM}
Same as in Fig.~\ref{fig:eta7}, but for MDM with $m = 9$ GeV/$c^2$.}
\end{figure}

As in Ref.~\cite{DelNobile:2013cva}, to determine the $\vmin$ interval corresponding to each detected energy interval $[\Ed_1, \Ed_2]$ for a particular experiment, we choose to use $90\%$ central quantile intervals of the modified response function, \ie we determine ${\vmin}_{,1}$ and ${\vmin}_{,2}$ such that the area under the function  $\vmin^{-r} \eR_{[\Ed_1, \Ed_2]}(\vmin)$ to the left of ${\vmin}_{,1}$ and the area to the right of ${\vmin}_{,2}$ are each separately $5\%$ of the total area. This gives the horizontal width of the crosses corresponding to the rate measurements in Figs.~\ref{fig:eta7}, \ref{fig:eta10}, \ref{fig:eta20} and \ref{fig:eta9MDM}. The horizontal placement of the vertical bar in the crosses corresponds to the maximum of the modified response function. The extension of the vertical bar, unless otherwise indicated, shows the $1 \sigma$ interval around the central value of the measured rate. In the figures, rather than drawing the new averages $\overline{\vmin^r \tilde{\eta}^i}$, we prefer to show $\vmin^{-r} \overline{\vmin^r \tilde{\eta}^{\,i}}$, so that a comparison can be easily made with the limits on $\tilde{\eta}^0$ described below, as well as with the previous literature on the spin-independent halo-independent method.

To compute upper bounds on $\tilde{\eta}^{\,0}$, the unmodulated part of $\tilde{\eta}$, from upper limits $R_{[\Ed_1, \Ed_2]}^{\rm lim}$ on the unmodulated rates, we follow the procedure proposed in Ref.~\cite{Fox:2010bz}. Because $\tilde{\eta}^{\,0}(\vmin)$ is by definition a non-decreasing function, the lowest possible $\tilde{\eta}^0(\vmin)$ function passing through a point $(v_0,\tilde{\eta}^{\,0})$ in $\vmin$ space is the downward step $\tilde{\eta}_0 \, \theta(v_0 - \vmin)$. The maximum value of $\tilde{\eta}_0$ allowed by a null experiment at a certain confidence level, denoted by $\tilde{\eta}^{\rm lim}(v_0)$, is determined by the experimental limit on the rate $R_{[\Ed_1, \Ed_2]}^{\rm lim}$ as
\beq
\tilde{\eta}^{\rm lim}(v_0) = \frac{R^{\rm lim}_{[\Ed_1, \Ed_2]}}{\int_0^{v_0} \ud \vmin \, \eR_{[\Ed_1, \Ed_2]}(\vmin)}
= \frac{R^{\rm lim}_{[\Ed_1, \Ed_2]}}{\eH_{[\Ed_1, \Ed_2]}(v_0)} \ .
\eeq

In Figs.~\ref{fig:eta7}, \ref{fig:eta10} and \ref{fig:eta20} we show the result of our halo-independent analysis for ADM with mass $m=7$ GeV/$c^2$, $m=10$ GeV/$c^2$ and $m=20$ GeV/$c^2$, respectively. These three masses are at the boundaries and at the center of the CDMS-II-Si region. We present $90\%$ CL bounds from the same experiments as in Fig.~\ref{fig:msigma} but added the CDMS modulation limit \cite{Ahmed:2012vq}. The crosses correspond to the DAMA (for $Q_{\rm Na}=0.30$) and CoGeNT 2014 modulation data (indicated by a subscript 1) and the unmodulated rate measurements of CoGeNT 2014 (indicated by a subscript 0) and CDMS-II-Si. For the CoGeNT 2014 modulation data points we show the modulus of the negative part of each cross with a thin blue line. In one of the four panels of each figure (the bottom left panel) we also include the CoGeNT 2011-2012 results (dashed crosses), for a comparison. In the four panels of each figure we present: modulation data and upper limits (top left), unmodulated rate measurements and upper limits (top right), unmodulated and modulated rate measurements and only the CDMS-II-Ge modulation limit (bottom left) and everything together (bottom right). The first three of these panels, less crowded than the last, include also the $\tilde{\eta}^{\,0}$ (dashed gray lines) and $\tilde{\eta}^{\,1}$ (dotted gray lines) functions corresponding to the SHM for $\sigma_{\rm ref}^A$ values $3\times 10^{-34}$ cm$^2$, $2\times 10^{-35}$ cm$^2$ and $3\times 10^{-36}$ cm$^2$ in Figs.~\ref{fig:eta7}, \ref{fig:eta10} and \ref{fig:eta20}, respectively, which are chosen within the CDMS-II-Si region in Fig.~\ref{fig:msigma}. For $m=7$ GeV/$c^2$, $\sigma_{\rm ref}^A=3\times10^{-34}~{\rm cm}^2$ is also within the CoGeNT 2014 region in Fig.~\ref{fig:msigma}, and in Fig.~\ref{fig:eta7} the corresponding $\tilde{\eta}^{\,1}$ seems to reproduce well the lowest energy bins of DAMA, while $\tilde{\eta}^{\,0}$ goes through the CDMS-II-Si data points. As it is expected from Fig.~\ref{fig:msigma}, the $\tilde{\eta}^{\,0}$ lines shown reproduce relatively well the CDMS-II-Si data points in Fig.~\ref{fig:eta10} but not in Fig.~\ref{fig:eta20}, while the $\tilde{\eta}^{\,1}$ lines do not reproduce well any of the data. However, in the three Figs.~\ref{fig:eta7}, \ref{fig:eta10} and \ref{fig:eta20} the $\tilde{\eta}^{\,0}$ lines are incompatible with XENON100, LUX and SuperCDMS limits.

Fig.~\ref{fig:eta9MDM} shows the same data points and limits of Fig.~\ref{fig:eta7} to \ref{fig:eta20} but for MDM with $m=9$ GeV/$c^2$, a mass for which the DAMA (with $Q_{\rm Na}=0.30$), CoGeNT and CDMS-II-Si regions overlap in Fig.~\ref{fig:msigma} (right panel). Qualitatively this figure is similar to Fig.~\ref{fig:eta10}, which corresponds to an ADM candidate very close in mass, but looking closely one can see differences in the bounds, due to the difference in cross sections.

In the halo-independent analysis the most constraining limits on ADM and MDM come from LUX, XENON100 and SuperCDMS while the CDMSlite bound is much above the DM-signal regions.

\section{Results and conclusions}

In the SHM analysis of the allowed regions and bounds in the $m$--$\sigma_{\rm ref}$ parameter space (Fig.~\ref{fig:msigma}), a combination of the 90\% CL LUX and CDMSlite limits or the new 90\% CL SuperCDMS limit by itself exclude the 90\% CL allowed regions of three experiments with a positive signal (DAMA, CoGeNT 2011-2012 modulation signal and 2014 unmodulated rate, and CDMS-II-Si) for WIMPs with anapole moment (ADM) and magnetic dipole moment (MDM) interactions. Although in our SHM analysis the DM-signal region is severely constrained by the CDMSlite limit, in our halo-independent analysis (Figs.~\ref{fig:eta7} to \ref{fig:eta9MDM}) this limit is much above the DM-signal region. The difference stems from the steepness of the SHM $\tilde{\eta}^{\,0}$ as a function of $\vmin$, which is constrained at low $\vmin$ by the CDMSlite and other limits.

In our halo-independent analysis, although the LUX bound is more constraining than the XENON100 limit, both cover the same range in $\vmin$ space and are limited to $\vmin\gtrsim600$ km/s for a WIMP mass of $7$ GeV/$c^2$, $450$ km/s for $10$ GeV/$c^2$ and $250$ km/s for $20$ GeV/$c^2$. This is due to the conservative suppression of the response function below $3.0$ keVnr assumed in this analysis for both LUX and XENON100 (see Ref.~\cite{DelNobile:2013gba} for details). Thus the LUX bound and the previous XENON100 bound exclude mostly the same data for ADM and MDM. In other words, almost all the DAMA, CoGeNT (both the 2011-2012 and 2014 data sets), and CDMS-II-Si energy bins that are not excluded by XENON100 are not excluded by LUX either. The situation remains of strong tension between the positive and negative results, as it was already before the LUX data.

At lower $\vmin$ values the most stringent bounds in our halo-independent analysis come from SuperCDMS and CDMSlite.

Even without considering the upper limits, in our halo-independent analysis there are problems in the DM signal regions by themselves: the crosses representing the unmodulated rate measurements of CDMS-II-Si are either overlapped or below the crosses indicating the modulation amplitude data as measured by CoGeNT (2011-2012 as well as 2014 data sets) and DAMA. This indicates strong tension between the CDMS-II-Si data on one side, and DAMA and CoGeNT modulation data on the other (these two seem largely compatible). 

\section*{Acknowledgments}

P.G.~was supported in part by NSF grant PHY-1068111. E.D.N., G.G.~and J.-H.H.~were supported in part by Department of Energy under Award Number DE-SC0009937. J.-H.H.~was also partially supported by Spanish Consolider-Ingenio MultiDark (CSD2009-00064).

\appendix

\section{ADM cross section}
\label{xsecADM}
In general, the differential cross section of the scattering of WIMP off a target nucleus is given by the {\it Golden Rule}, 
\begin{equation}
\label{eq:goldenrule}
{\rm d}\sigma_T
=\frac{2\pi}{v}\delta(E_{{\bf p}'}+\epsilon_{{\bf k}'}-E_{{\bf p}}-\epsilon_{{\bf k}})\overline{\left|T_{\rm fi}\right|^2}\frac{{\rm d}{\bf k'}}{(2\pi)^3}\ ,
\end{equation}
where $v$ is the modulus of the relative velocity $\bfv$ of the WIMP with respect to the nucleus, $E_{{\bf p}'}$, $E_{{\bf p}}$ and $\epsilon_{{\bf k}'}$, $\epsilon_{{\bf k}}$ are the final and initial energies of the nucleus and the WIMP respectively, ${\bf p}'$, ${\bf p}$ and ${\bf k}'$,${\bf k}$ are the final and initial momenta of the nucleus and the WIMP, respectively, and an overline denotes an average over the initial and sum over the final spin polarizations. Here, the transition amplitude $T_{\rm fi}$ at the leading order is defined by
\begin{equation}
\label{eq:Tfi}
S_{\rm fi}=i\int{\rm d}^4x\langle {\bf k}',s',\lambda'|\hat{\Lag}_I(x)|{\bf k},s,\lambda\rangle\equiv i(2\pi)\delta(E_{{\bf p}'}+\epsilon_{{\bf k}'}-E_{{\bf p}}-\epsilon_{{\bf k}})T_{\rm fi} \ ,
\end{equation}
where $S_{\rm fi}$ is the scattering matrix element between the initial and final states of the WIMP-nucleus system, $|{\bf k},s,\lambda\rangle$ and $|{\bf k},s',\lambda'\rangle$ with the initial (final) spin polarizations $s$ ($s'$) and $\lambda$ ($\lambda'$) of the WIMP and the nucleus, respectively, and $\hat \Lag_I(x)$ is the interaction lagrangian.
We use the state normalization
\begin{equation}
\label{eq:statenorm}
\langle{\bf k'},s',\lambda'|{\bf k},s,\lambda\rangle
=\delta_{ss'}\delta_{\lambda'\lambda}(2\pi)^3\delta^{(3)}({\bf k}'-{\bf k})\ ,
\end{equation}
which is compatible with Eqs.~\eqref{eq:goldenrule} and \eqref{eq:Tfi}, and the conservation of the total momentum.

Integrating the ADM interaction lagrangian $\frac{1}{2} \frac{g}{\Lambda^2} \, \bar{\chi} \gamma^{\mu} \gamma^5 \chi \, \partial^\nu \hat F_{\mu\nu}$ by parts, we can write an equivalent lagrangian $\hat{\Lag}_{\rm I}(x)=-\hat j_\mu(x)\hat A^\mu(x)$ in terms of the electromagnetic current of the ADM Majorana fermion (notice that $\hat{}$ denotes operators unless otherwise stated)
\begin{equation}
\hat j^\mu(x)=-\frac{g}{2\Lambda^2}(g^{\mu\lambda}\partial^\nu\partial_\nu -\partial^\mu\partial^\lambda)\bar\chi(x)\gamma_{\lambda}\gamma_5\chi(x) .
\end{equation}
With this expression for $\hat\Lag_I(x)$, the right-hand side of Eq.~\eqref{eq:Tfi} can be written as
\begin{eqnarray}
\label{eq:RHS}
{\rm (RHS)}
&=&-i\int{\rm d}^4x
\langle{\bf k}',s'|\hat j^\mu(x)|{\bf k},s\rangle
\langle\lambda'|\hat A_\mu(x)|\lambda\rangle\nonumber\\
&=&-i\langle{\bf k}',s'|\hat j^\mu(0)|{\bf k},s\rangle
\int{\rm d}^4x
e^{-iq\cdot x}
\langle\lambda'|\hat A_\mu(x)|\lambda\rangle \nonumber\\
&=&-i\langle{\bf k}',s'|\hat j^\mu(0)|{\bf k},s\rangle
\langle\lambda'|\hat A_\mu(q)|\lambda\rangle \ ,
\end{eqnarray}
where $\hat A^\mu(q)$ is the four-dimensional Fourier transform of $\hat A^\mu(x)$, the four-momentum transfer $q^\mu=k^\mu-k'^\mu$ with $k^\mu=(\epsilon_{\bf k},{\bf k})$ and $k'^\mu=(\epsilon_{{\bf k}'},{\bf k}')$, and
$|{\bf k},s,\lambda\rangle=|{\bf k},s\rangle|\lambda\rangle$ with normalization $\langle{\bf k}',s'|{\bf k},s\rangle=(2\pi)^3\delta_{s's}\delta^{(3)}({\bf k}'-{\bf k})$. Notice that $|\lambda\rangle$ and $|\lambda'\rangle$ have an implicit dependence on the nucleus momenta (determined by the ADM momenta, ${\bf k}'$ and ${\bf k}$).

The matrix element of the ADM electromagnetic current at the origin between ADM momentum and spin eigenstates can be written as
\begin{equation}
j^\mu_{s's}(0)=(\rho^\chi_{ss'}(0),\bol{j}^\chi_{ss'}(0))=q^2\left(g^{\mu\lambda}-\frac{q^\mu q^\lambda}{q^2}\right)
\frac{g}{\Lambda^2}
\frac{1}{\sqrt{2\epsilon_{\bf k} 2\epsilon_{{\bf k}'}}}
\bar{\rm u}_{{\bf k}'s'}\gamma_\lambda\gamma_5 {\rm u}_{{\bf k}s}\ ,
\end{equation}
where ${\rm u}_{{\bf k}s}$ (${\rm u}_{{\bf k}'s'}$) is the Majorana spinor wave function with momentum ${\bf k}$ (${\bf k}'$) and polarization $s$ ($s'$), normalized as $\bar u_{{\bf k}s'}u_{{\bf k}s}=2m\delta_{s's}$.
In the non-relativistic limit,
$\bar{\rm u}_{{\bf k}'s'}\gamma^\mu\gamma_5{\rm u}_{{\bf k}s}
=\sqrt{2\epsilon_{\bf k} 2\epsilon_{{\bf k}'}}\left(\frac{{\bf k}+{\bf k}'}{2m}\cdot{\bf s},{\bf s}\right)$,
where ${\bf s}=\chi_{s'}^\dagger\bol{\sigma}\chi_s$ with the vector of Pauli matrices $\bol{\sigma}=(\sigma_1,\sigma_2,\sigma_3)$ and the non-relativistic Pauli spinor $\chi_s$ for the spin polarization $s$.
A simple algebraic calculation shows that the non-relativistic limit of $j^\mu_{s's}(0)$ is
\begin{eqnarray}
\label{eq:jADM1}
\rho^\chi_{s's}(0)&=&{\bf q}^2\frac{g}{\Lambda^2}
{\bf s}\cdot\frac{{\bf k}+{\bf k}'}{2m} \ , \\
\label{eq:jADM2}
\bol{j}^\chi_{s's}(0)&=&{\bf q}^2\frac{g}{\Lambda^2}
{\bf s}_{\rm T} \ ,
\end{eqnarray}
where ${\bf s}_{\rm T}={\bf s}-({\bf s}\cdot\hat{\bf q})\hat{\bf q}$ with $\hat{\bf q}={\bf q}/|{\bf q}|$ is the component of ${\bf s}$ transverse to ${\bf q}={\bf k}-{\bf k}'$.

As a solution to the Heisenberg equation of motion in the Lorenz gauge $\partial_\mu \hat A^\mu(x)=0$, the four-dimensional Fourier transform $\hat A^\mu(q)$ of $\hat A^\mu(x)$ generated by the nucleus can be written as
\begin{equation}
\hat A^\mu(q)=-\frac{e \hat J^\mu(q)}{q^2}\ ,
\end{equation}
in terms of the four-dimensional Fourier transform $\hat J^\mu(q)$ of the nucleus electromagnetic current operator $\hat J^\mu(x)$ ($e$ is the electromagnetic coupling constant). Knowing that $|\lambda\rangle$ and $|\lambda'\rangle$ are energy eigenstates with eigenvalues $E_{{\bf p}}$ and $E_{{\bf p}'}$, respectively, the matrix element $\langle\lambda'|\hat A_\mu(q)|\lambda\rangle$ can be simply written as
\begin{equation}
\label{eq:JtoA}
\langle\lambda'|\hat A_\mu(q)|\lambda\rangle
=2\pi\delta(E_{{\bf p}'}+\epsilon_{{\bf k}'}-E_{{\bf p}}-\epsilon_{{\bf k}})\frac{e\langle\lambda'|\hat J^\mu({\bf q})|\lambda\rangle}{|{\bf q}|^2}\ ,
\end{equation}
where $\hat J^\mu({\bf q})$ is the three-dimensional Fourier transform of $\hat J^\mu(x^0=0,{\bf x})$, and $q^2=-|{\bf q}|^2$ in the non-relativistic limit.

By inserting Eqs.~\eqref{eq:jADM1}, \eqref{eq:jADM2} and \eqref{eq:JtoA} into Eq.~\eqref{eq:RHS}, we can write the transition amplitude $T_{\rm fi}$ as
\begin{equation}
\label{eq:Tficm}
T_{\rm fi}=\frac{e_p g}{\Lambda^2}{\bf s}_{\rm T}\cdot
\left[\frac{{\bf k}+{\bf k}'}{2m}\rho_{\lambda'\lambda}({\bf q})-{\bf J}_{\lambda'\lambda}({\bf q})\right]\ ,
\end{equation}
in terms of the nuclear current matrix element
$\langle\lambda'|\hat J^\mu({\bf q})|\lambda\rangle
=(\rho_{\lambda'\lambda}({\bf q}),\bol{J}_{\lambda'\lambda}({\bf q}))$.
Notice that we have replaced ${\bf s}$ in the $\rho_{\lambda'\lambda}({\bf q})$ term with ${\bf s}_{\rm T}$ since for an elastic scattering ${\bf q}\cdot({\bf k}+{\bf k}')=({\bf k}-{\bf k}')\cdot({\bf k}+{\bf k}')=0$.

To write the nuclear charge and current density matrix elements $\rho_{\lambda'\lambda}({\bf q})$ and $\bol{J}_{\lambda'\lambda}({\bf q})$ in terms of the quantities defined in de Forest and Walecka \cite{deForest:1966}, we decompose $J^\mu(q)$ in the reference frame defined as the target nucleus rest frame with the $z$ axis in the direction of the momentum exchange ${\bf q}_{\rm lab}$, and the $x$ and $y$ axes in the plane transverse to ${\bf q}_{\rm lab}$. Spherical basis vectors ${\bf e}_{\pm1}=(\hat{\bf x} \pm i\hat{\bf y})/\sqrt{2}$ are also introduced. In terms of four-vectors, the de Forest and Walecka frame can be defined by the four mutually orthogonal four-vectors $p^\mu$, $q^\mu-(p\cdot q/m_T^2)p^\mu$, $e_{+1}^{\mu}$ and $e_{-1}^{\mu}$, which in this frame read $p^\mu=(m_T,{\bf 0})$, $q^\mu-(p\cdot q)p^\mu/m_T^2=(0,{\bf q}_{\rm lab})$, $e_{+1}^{\mu}=(0,{\bf e}_{+1})$, and $e_{-1}^{\mu}=(0,{\bf e}_{-1})$. A short algebraic calculation reveals that a four-vector $J^\mu(q)$ that obeys the conservation law $q\cdot J=0$ can be written as
\begin{equation}
J^\mu=\frac{-q^2}{{\bf q}_{\rm lab}^2}\frac{\rho^{\rm lab}}{m_T}
\left(p^\mu-\frac{p\cdot q}{q^2}q^\mu\right)
+J_{\rm lab,+1}e_{+1}^{\mu}+J_{\rm lab,-1}e_{-1}^{\mu}\ ,
\end{equation}
where $\rho^{\rm lab}=J\cdot p/m_T$ and $J^{{\rm lab}}_{\alpha}=-J\cdot e^*_{\alpha}$ ($\alpha=\pm1$) are the time and spherical components of $J^\mu$, respectively, in the de Forest and Walecka frame. Using the fact that $p^\mu-(p\cdot q/q^2)q^\mu=(p^\mu+p'^\mu)/2$ and $q^2=-{\bf q}_{\rm lab}^2$ in the non-relativistic limit, we find
\begin{eqnarray}
\label{eq:Jcm}
\rho_{\lambda'\lambda}({\bf q})&=&\rho^{\rm lab}_{\lambda'\lambda}({\bf q})+O(v^2)\ ,\nonumber\\
\bol{J}_{\lambda'\lambda}({\bf q})&=&\rho^{\rm lab}_{\lambda'\lambda}({\bf q})\frac{{\bf p}+{\bf p}'}{2m_T}+\bol{J}^{T,{\rm lab}}_{\lambda'\lambda}({\bf q})+O(v^2),
\end{eqnarray}
where ${\bf p}$ and ${\bf p}'$ are the initial and final momenta of the nucleus in a given frame, and $\bol{J}^{T,{\rm lab}}_{\lambda'\lambda}({\bf q})
=\bol{J}^{{\rm lab}}_{{\lambda'\lambda},+1}\cdot{\bf e}_{+1}
+\bol{J}^{{\rm lab}}_{{\lambda'\lambda},-1}\cdot{\bf e}_{-1}$.
Inserting these equations into Eq.~\eqref{eq:Tficm},
\begin{equation}
T_{\rm fi}=\frac{e g}{\Lambda^2}{\bf s}_{\rm T}\cdot[{\bf V}_{\rm T}\rho^{\rm lab}_{\lambda'\lambda}({\bf q}) - \bol{J}^{T,{\rm lab}}_{\lambda'\lambda}({\bf q})],
\end{equation}
where ${\bf V}_{\rm T}$ is the transverse velocity, ${\bf V}_{\rm T}=({\bf k}+{\bf k}')/2m-({\bf p}+{\bf p}')/2m_T$. Notice that ${\bf V}_{\rm T}\cdot{\bf q}=0$ as the name indicates.
Using ${\bf k}'={\bf k}-{\bf q}$, ${\bf p}'={\bf p}+{\bf q}$, and $\bfv={\bf k}/m-{\bf p}/m_T$, the transverse velocity can also be written as
\begin{equation}
{\bf V}_{\rm T}=\frac{{\bf k}}{m}-\frac{{\bf p}}{m_T}-\frac{{\bf q}}{2m}-\frac{{\bf q}}{2m_T}
=\bfv-\frac{{\bf q}}{2\mu_T}.
\end{equation}
This equation combined with the transversality condition of ${\bf V}_{\rm T}$, ${\bf V}_{\rm T}\cdot{\bf q}=0$, gives the relation
\begin{equation}
\label{eq:VT2}
v^2={\bf V}_{\rm T}^2+\frac{{\bf q}^2}{4\mu_T^2} \ ,
\end{equation}
which will be used later.

To compute $\overline{|T_{\rm fi}|^2}$ we need the spin sums and averages of the square of the nuclear charge and the current densities. In terms of the longitudinal and transverse electromagnetic form factors $F_{\rm L}^2({\bf q})$ and $F_{\rm T}^2({\bf q})$ in Refs.~\cite{Donnelly:1975ze,Donnelly:1984rg,deForest:1966}, we have
\begin{eqnarray}
\overline{\rho^{\rm lab}({\bf q})\rho^{{\rm lab}*}({\bf q})}&=&4\pi F_{\rm L}^2({\bf q}^2)\ ,\\
\overline{\rho^{\rm lab}({\bf q})J^{T,{\rm lab},\alpha*}({\bf q})}&=&0 \ ,\\
\overline{J^{T,{\rm lab},\alpha}({\bf q})J^{T,{\rm lab},\alpha'*}({\bf q})}&=&2\pi F_{\rm T}^2({\bf q}^2)\delta_{\alpha\alpha'}\ ,
\end{eqnarray}
so
\begin{equation}
\overline{|T_{\rm fi}|^2}=\frac{e^2g^2}{\Lambda^4}
\left[\overline{({\bf s}_{\rm T}\cdot{\bf V}_{\rm T})({{\bf s}_{\rm T}}^*\cdot{\bf V}_{\rm T})}4\pi F_{\rm L}^2({\bf q}^2)+\overline{{\bf s}_{\rm T}\cdot{{\bf s}_{\rm T}}^*}2\pi F_{\rm T}^2({\bf q}^2)\right].
\end{equation}
The remaining spin sums and averages of the ADM spin polarizations are evaluated 
using $\overline{s_i {s_j}^*}={\rm tr}(\sigma_i\sigma_j)/2=\delta_{ij}$, from which $\overline{s_{T,i} {s_{T,j}}^*}=\overline{(s_i-{\bf s}\cdot\hat {\bf q}\hat q_i)(s_j-{\bf s}\cdot\hat {\bf q}\hat q_j)^*}=\delta_{ij}-\hat q_i\hat q_j$. Hence
\begin{equation}
\label{eq:Tfi2}
\overline{|T_{\rm fi}|^2}=\frac{4\pi e^2g^2}{\Lambda^4}
\left[ {\bf V}_{\rm T}^2F_{\rm L}^2({\bf q}^2) + F_{\rm T}^2({\bf q}^2)\right].
\end{equation}
Notice that this expression is frame independent.

By using the fact that, in the center of mass frame, $\delta(E_{{\bf p}'}+\epsilon_{{\bf k}'}-E_{{\bf p}}-\epsilon_{{\bf k}}){\rm d}|{\bf k}'|=1/(|{\bf k}'|/M+|{\bf k}'|/m)=\mu_T/|{\bf k}'|$ and $\mu_T \bfv=\bf k'$, we get
\begin{equation}
\label{eq:QMxsec}
\frac{{\rm d}\sigma_T}{{\rm d}\Omega_{\rm cm}}
=\frac{\mu_T^2}{4\pi^2}\overline{\left|T_{\rm fi}\right|^2}.
\end{equation}
Using the relation
$E_R=(\mu_T^2v^2/m_T)(1-\cos\theta_{\rm cm})$, the differential cross section in terms of the energy transfer $E_R$ can be obtained as
\begin{equation}
\label{eq:xsecER}
\frac{{\rm d}\sigma_T}{{\rm d}E_R}=\frac{m_T}{2\pi v^2}\overline{|T_{\rm fi}|^2},
\end{equation}
which is frame independent.
Finally, by using Eq.~\eqref{eq:VT2}, the cross section follows from Eqs.~\eqref{eq:Tfi2} and \eqref{eq:xsecER} as
\begin{equation}
\frac{{\rm d}\sigma}{{\rm d}E_R}=\frac{8\pi m_T}{v^2}\frac{\alpha g^2}{\Lambda^2}\left[\left(v^2-\frac{{\bf q}^2}{4\mu_T^2}\right)F_{\rm L}^2({\bf q}^2)+F_{\rm T}^2({\bf q}^2)\right]\ ,
\end{equation}
where $\alpha=e^2/4\pi$ is the fine structure constant. Inserting Eqs.~\eqref{eq:FE} and \eqref{eq:FM} into this equation we obtain
\begin{equation}
\label{eq:ADMxsecNR}
\frac{d\sigma_T}{dE_R}=\sigma^A_{\rm ref}\frac{m_T}{\mu_N^2}
\frac{1}{v^2}\left[Z^2\left(v^2-\frac{{\bf q}^2}{4\mu_T^2}\right)F^2_{{\rm E},T}({\bf q}^2)
+\frac{\lambda_T^2}{\lambda_N^2}\frac{m_T^2}{m_N^2}\left(\frac{J_T+1}{3 J_T}\right)
\frac{{\bf q}^2}{2m_T^2}F_{{\rm M},T}({\bf q}^2)
\right].
\end{equation}

We can easily test this equation using the more familiar field theoretic formulation for $J_T=1/2$. For this case we can represent the longitudinal and
transverse form factors, $F_{\rm L}({\bf q}^2)$ and $F_{\rm T}({\bf q}^2)$, in terms of familiar electromagnetic structure functions $F_1(q^2)$ and $F_2(q^2)$, also called the Dirac and Pauli form factors, respectively. They are defined by
\begin{equation}
\langle{\bf p}',\lambda'|\hat J^\mu(x)|{\bf p},\lambda\rangle
=\frac{1}{\sqrt{2E_{{\bf p}'}2E_{{\bf p}}}}\bar u_{T,\lambda'}({\bf p}')\left[F_1(q^2)\gamma^\mu+\frac{i}{2m_T}F_2(q^2)\sigma^{\mu\nu}q_\nu\right]u_{T,\lambda}({\bf p})e^{i(p'-p)x}\ ,
\end{equation}
where $|{\bf p},\lambda\rangle$ is the nuclear state with momentum ${\bf p}$ and spin polarization $\lambda$, normalized as $\langle{\bf p}',\lambda'|{\bf p},\lambda\rangle=\delta_{\lambda'\lambda}(2\pi)^3\delta^{(3)}({\bf p}'-{\bf p})$, and $u_T$ is the Dirac spinor wave function of the target nucleus. Notice that the state $|{\bf p},\lambda\rangle$ differs from $|\lambda\rangle$ used above (in particular in Eq.~\eqref{eq:statenorm}) by a divergent constant factor, $|{\bf p},\lambda\rangle=\sqrt{(2\pi)^3\delta^{(3)}({\bf 0})}|\lambda\rangle$.

Since in the non-relativistic limit
\begin{eqnarray}
\bar u_{T,\lambda'}({\bf p}')\gamma^\mu u_{T,\lambda}({\bf p})
&=& \left(2m_T\xi^\dagger_{\lambda'}\xi_{\lambda},~-i\xi^\dagger_{\lambda'}{\bf q}\times\bol{\sigma}\xi_{\lambda}\right)\ ,\\
\bar u_{T,\lambda'}({\bf p}')\frac{i\sigma^{\mu\nu}q_\nu}{2m_T} u_{T,\lambda}({\bf p})
&=& \left(-\frac{{\bf q}^2}{2m_T}\xi^\dagger_{\lambda'}\xi_{\lambda},~-i\xi^\dagger_{\lambda'}{\bf q}\times\bol{\sigma}\xi_{\lambda}\right)\ ,
\end{eqnarray}
where $\xi_\lambda$ is the non-relativistic Pauli spinor, 
and $F_1(q^2=0)=Z$ and $F_2(q^2=0)=(g_T-2Z)/2$ with the target nucleus $g$-factor $g_T$, the matrix element of $\hat J^{\mu}({\bf q})=\int d{\bf x} \hat J^{\mu}(0,{\bf x})e^{i{\bf q}\cdot{\bf x}}$ can be written as
\begin{equation}
\langle{\bf p}',\lambda'|\hat J^{\mu}({\bf q})|{\bf p},\lambda\rangle
=(2\pi)^3\delta^{(3)}({\bf 0})
\left(G_{\rm E}({\bf q}^2)\xi_{\lambda'}^\dagger\xi_\lambda,~-i\frac{G_{\rm M}({\bf q}^2)}{2m_T}\xi_{\lambda'}^\dagger{\bf q}\times\bol{\sigma}\xi_\lambda\right),
\end{equation}
where $G_{\rm E}({\bf q}^2)=F_1({\bf q}^2)-({\bf q}^2/4m_T^2)F_2({\bf q}^2)$ and $G_{\rm M}({\bf q}^2)=F_1({\bf q}^2)+F_2({\bf q}^2)$ are the electric and magnetic Sachs form factors, respectively. From this we can read off
\begin{eqnarray}
\rho_{\lambda'\lambda}({\bf q})&=&G_{\rm E}({\bf q}^2)\xi_{\lambda'}^\dagger\xi_\lambda~,\\
\bol{J}_{\lambda'\lambda}({\bf q})&=&-i\frac{G_{\rm M}({\bf q}^2)}{2m_T}\xi_{\lambda'}^\dagger{\bf q}\times\bol{\sigma}\xi_\lambda \ .
\end{eqnarray} 

By summing/averaging over spin polarizations, we get
\begin{eqnarray}
\overline{\rho({\bf q})\rho({\bf q})}
&=&4\pi F_{\rm L}^2({\bf q})
=G_{\rm E}^2({\bf q}^2)\ ,
\\
\overline{J^{T,\alpha}({\bf q})J^{T,\alpha'*}({\bf q})}
&=&\delta_{\alpha\alpha'}2\pi F_{\rm T}^2({\bf q})
=\delta_{\alpha'\alpha}\frac{{\bf q}^2}{4m_T^2}G_{\rm T}^2({\bf q}^2)~,
\end{eqnarray}
and we can identify the longitudinal and transverse form factors as
\begin{equation}
F_{\rm L}^2({\bf q}^2)=\frac{1}{4\pi}G_{\rm E}^2({\bf q}^2)~ {\rm and}~
F_{\rm T}^2({\bf q}^2)=\frac{{\bf q}^2}{8\pi m_T^2}G_{\rm T}^2({\bf q}^2).
\end{equation}

With this identification one can easily check that the non-relativistic limit of the fully relativistic differential cross section for a nucleus with spin $1/2$ is that in Eq.~\eqref{eq:ADMxsecNR} with $J_T=1/2$.

\end{document}